\documentclass[12pt]{article}

\pdfoutput=1

\usepackage{graphicx}
\usepackage{epsfig}

\usepackage{amsfonts}
\usepackage{amssymb}
\usepackage{amsmath}
\usepackage{dsfont}
\usepackage{pifont}
\usepackage{bbm}
\usepackage{multirow}
\usepackage{latexsym}
\usepackage{verbatim}
\usepackage{mcite}
\usepackage{cite}
\usepackage{units}
\usepackage[all]{xy}
\usepackage{arydshln}
\usepackage{cancel}
\usepackage{slashed}

\def\beq{\begin{equation}}
\def\eeq{\end{equation}}
\def\beqa{\begin{eqnarray}}
\def\eeqa{\end{eqnarray}}

\def\bfone{\relax{\rm 1\kern-.35em 1}}

\newcommand{\be}{\begin{equation}}
\newcommand{\ee}{\end{equation}}
\newcommand{\ben}{\begin{displaymath}}
\newcommand{\een}{\end{displaymath}}
\newcommand{\bea}{\begin{eqnarray}}
\newcommand{\eea}{\end{eqnarray}}

\newcommand{\bean}{\begin{eqnarray*}}
\newcommand{\eean}{\end{eqnarray*}}

\DeclareMathAlphabet{\mathpzc}{OT1}{pzc}{m}{it}

\begin{document}
\pagestyle{plain}
\makeatletter \@addtoreset{equation}{section} \makeatother
\renewcommand{\thesection}{\arabic{section}}
\renewcommand{\theequation}{\thesection.\arabic{equation}}
\renewcommand{\thefootnote}{\arabic{footnote}}
\setcounter{page}{1} \setcounter{footnote}{0}

\begin{titlepage}

\begin{flushright}
\vspace{1cm}
\end{flushright}

\begin{center}

\bigskip

\vskip 0.5cm
{\LARGE \bf A second look at gauged supergravities \\[4mm] from fluxes in M-theory} \\[6mm]
\vskip 0.5cm
{\bf Jean-Pierre Derendinger \, and \,  Adolfo Guarino}\\[0mm]
\let\thefootnote\relax\footnote{{\tt derendinger@itp.unibe.ch, guarino@itp.unibe.ch}}\\
\vskip 0.5cm
{\em
Albert Einstein Center for Fundamental Physics, Institute for Theoretical Physics, \\ Bern University, Sidlerstrasse 5, CH–3012 Bern, Switzerland}
\vskip 0.8cm
\end{center}
\vskip 1cm
\begin{center}
{\bf ABSTRACT}\\[3ex]
\begin{minipage}{13cm}
\small

We investigate reductions of M-theory beyond twisted tori by allowing the presence of KK6 monopoles (KKO6-planes) compatible with $\,\mathcal{N}=4\,$ \mbox{supersymmetry} in four dimensions. The presence of KKO6-planes proves \mbox{crucial} to achieve full moduli stabilisation as they generate new universal \mbox{moduli} powers in the scalar potential. The resulting gauged supergravities turn out to be compatible with a weak $\,G_{2}\,$ holonomy at $\,\mathcal{N}=1\,$ as well as at some non-supersymmetric AdS$_{4}$ vacua. The \mbox{M-theory} flux vacua we present here cannot be obtained from ordinary type IIA orientifold reductions \mbox{including} background fluxes, \mbox{D6-branes} (O6-planes) and/or KK5 (KKO5) sources. However, from a four-dimensional point of view, they still admit a \mbox{description} in terms of so-called non-geometric fluxes. In this sense we provide the \mbox{M-theory} interpretation for such non-geometric type IIA flux vacua.

\end{minipage}
\end{center}
\vfill
\end{titlepage}


\tableofcontents

\newpage

\section{Motivation}

How to get masses from extra dimensions \cite{Scherk:1979zr} has captured the attention of theoretical physicists during the last thirty five years. How massless theories in higher dimensions lead to massive theories in lower dimensions remains at the core of the connection between strings or M-theory and the real world. One may think of two approaches. The first is the \textit{top-down} approach where a higher-dimensional theory like strings or M-theory is selected and then a lower-dimensional effective model is derived from the choice of a compactification scheme. In this way the dynamics in lower dimensions follows from the reduction prescription. The higher-dimensional interpretation of such effective models is clear but, as a downside, one often engineers classes of compactifications which do not produce satisfactory physics. Alternatively, the \textit{bottom-up} approach begins with an effective field theory (EFT) in lower dimensions selected using low-energy dynamical or phenomenological criteria. Only then can one try to relate such well-motivated models to more fundamental theories in higher dimensions like strings or M-theory. This may be feasible if some guiding principles are respected in the selection of the EFT.

A crucial ingredient in the construction of an EFT expected to describe classes of strings/M-theory compactifications is the number of preserved or broken supersymmetries. This is even more relevant than the space-time dimension since the existence of supercharges severely restricts the field content and the structure of the effective action in all dimensions. For the cases of $32$ (maximal) and $16$ (half-maximal) supercharges in four dimensions (4d), the guiding principle which governs the structure of the EFT is the embedding tensor (ET) formalism. This framework allows for a systematic exploration of $\,\mathcal{N}=8\,$ (maximal) \cite{deWit:2007mt} and $\,\mathcal{N}=4\,$ (half-maximal) \cite{Schon:2006kz} effective supergravity models -- in the form of gauged supergravities -- which, on the other hand, are of special interest due to their plausible realisation in higher dimensions as maximally supersymmetric and $1/2$-BPS backgrounds. 

However the identification between parameters in the embedding tensor formalism and quantities in a higher-dimensional theory turns out to be a  subtle task and has occasionally led to some confusion in the literature. This has been for instance the case for the effective STU-models of ref.~\cite{Derendinger:2004jn} arising from \textit{massive} type IIA orientifold reductions including background fluxes, D6-branes and O6-planes. These were the first string constructions featuring \textit{full} moduli stabilisation in a vacuum without requiring non-perturbative effects, such as Euclidean brane instantons or gaugino condensation, to stabilise the K\"ahler moduli \cite{Kachru:2003aw}. In ref.~\cite{Derendinger:2004jn} an $\,\mathcal{N}=1\,$ flux-induced superpotential $W_{\textrm{IIA}}$ was presented and the fluxes (couplings in $W_{\textrm{IIA}}$) were related to $\,\mathcal{N}=4\,$ gauging parameters, thus establishing a  correspondence between flux compactifications and ($\mathcal{N}=1$ truncations of) $\mathcal{N}=4$ gauged supergravity in the context of type IIA orientifold reductions (16 supercharges). The string vacuum of ref.~\cite{Derendinger:2004jn} was reconsidered in ref.~\cite{Villadoro:2007yq} and found to actually require the presence of KK5 monopoles due to a relation of the form \cite{Villadoro:2007yq,Andriot:2014uda}
\beq
\label{ww_IIA}
\begin{array}{ccc}
\omega \, \omega \neq 0 &\Rightarrow &  \textrm{Net charge of KK5 (KKO5) sources}
\end{array}
\eeq
involving the Scherk-Schwarz metric $\omega$-flux along the six-dimensional internal space $X_{6}$. This result indicated the necessity to extend the twisted tori picture of ref.~\cite{Scherk:1979zr}, which demands $\,\omega \, \omega = 0\,$ as a consistency relation. However, and only after the advent of the embedding tensor formalism, a thorough study of type IIA orientifold reductions \cite{Dall'Agata:2009gv} showed that $\omega \, \omega \neq 0$ violates the consistency conditions of $\,\mathcal{N}=4\,$ gauged supergravity \cite{Schon:2006kz}. As a consequence, the string vacuum of ref.~\cite{Derendinger:2004jn} is not a solution of $\,\mathcal{N}=4\,$ gauged supergravity although it still is a perfectly acceptable solution of the $\mathcal{N}=1$ supergravity specified by the superpotential $W_{\textrm{IIA}}$. Nevertheless various type IIA orientifold models actually corresponding to $\,\mathcal{N}=4\,$ gauged supergravities, \textit{i.e.} satisfying $\,\omega \, \omega = 0\,$, have been worked out afterwards on the basis of the ET formalism \cite{Dall'Agata:2009gv,Dibitetto:2011gm}. In all the cases where full moduli stabilisation occurred,  the \textit{massive} version \cite{Romans:1985tz} of the type IIA theory was needed.

Gauged supergravities related to M-theory reductions to four dimensions have been much less explored 
\cite{House:2004pm,DallAgata:2005fm,Hull:2006tp,Looyestijn:2010pb} than their type IIA relatives\footnote{Consistent truncations of M-theory beyond the toroidal setup we discuss in this work have been discussed in refs~\cite{Castellani:1983tc,Castellani:1983yg,Cassani:2011fu,Cassani:2012pj}.}. Ref.~\cite{DallAgata:2005fm} investigated in detail Scherk-Schwarz reductions on $G_{2}$-manifolds in the presence of background fluxes, derived an $\,\mathcal{N}=1\,$ flux-induced superpotential $\,W_{\textrm{M-theory}}\,$ and established the connection to the previous type IIA orientifold constructions by exploiting their underlying SU(3)-structure. The resulting STU-models corresponded to ($\mathcal{N}=1$ truncations of) $\,\mathcal{N}=8\,$ gauged supergravities incompatible with full moduli stabilisation. Remarkably the authors identified a mismatch\footnote{See also ref.~\cite{Aldazabal:2006up}.} between the $\,\mathcal{N}=1\,$ superpotentials of the M-theory models \mbox{(32 supercharges)} and of the type IIA orientifold models \mbox{(16 supercharges)} which can be summarised as
\beq
\label{Superpotential_comparison_ISO_Intro}
W_{\textrm{M-theory}}\,\,=\,\, \left. W_{\textrm{IIA}} \right|_{a_{3}=0}  \,\,- 3 \, c_{3}'\, T^{2} - 3 \, d_{0} \, S  \,T  \ ,
\eeq
where $\,a_{3}\,$ is the Romans mass\footnote{The IIA Romans mass parameter \cite{Romans:1985tz} is not generated upon (non-singular \cite{Hull:1998vy}) ordinary reductions of M-theory.} and the flux parameters $\,(c_{3}' \,,\, d_{0})\,$ are metric $\omega$-fluxes in M-theory with no counterpart in the standard type IIA orientifold constructions\footnote{They would correspond to \textit{non-geometric} fluxes \cite{Shelton:2005cf,Aldazabal:2006up,Dibitetto:2011gm} in a modern approach to type IIA flux compactifications.}. For this reason, they were set to zero in ref.~\cite{DallAgata:2005fm} in order to have a neat $\,\textrm{SU}(3) \subset \textrm{G}_{2}\,$ embedding of the internal manifolds (6d \textit{vs} 7d) underlying the type IIA orientifold and the M-theory reductions. In this work we will investigate several aspects of these genuine M-theory fluxes. 

One of our main results is that \textit{full} moduli stabilisation can be achieved in M-theory scenarios provided that the fluxes $\,(c_{3}' \,,\, d_{0})\,$ are activated. The minimally setup requires an $\,\mathcal{N}=8 \rightarrow \mathcal{N}=4\,$ breaking of supersymmetries (from $32$ supercharges to $16$) in the effective STU-models. Using the embedding tensor formalism as an organising principle -- for this we will derive a precise ET/flux dictionary in M-theory -- we will show that the set of $\,\mathcal{N}=4\,$ consistency relations is \textit{compatible} with a relaxation of the Scherk-Schwarz conditions $\,\omega \, \omega = 0\,$ involving the metric $\omega$-flux in M-theory, in contrast to what happened in the type IIA case. Along the lines of ref.~\cite{Villadoro:2007yq}, we will introduce the corresponding KK6 monopoles entering the relation
\beq
\label{ww_IM-theory}
\begin{array}{ccc}
\omega \, \omega \neq 0 &\Rightarrow &  \textrm{Net charge of KK6 (KKO6) sources} \ ,
\end{array}
\eeq
which now involves the Scherk-Schwarz metric $\omega$-flux along the seven-dimensional internal space $X_{7}$, and discuss their compatibility with preserving $\,\mathcal{N}=4\,$ supersymmetry in the effective action. The aim of this work is to extend the study of type IIA/M-theory reductions initiated in ref.~\cite{DallAgata:2005fm} by exploiting the power of the embedding tensor formalism used to systematically analyse maximal and half-maximal gauged supergravitites in four dimensions.

The paper is organised as follows. In section~\ref{sec:M-theoryG2} we review the reductions of M-theory on \mbox{$G_{2}$-manifolds} with fluxes \cite{DallAgata:2005fm} and their interpretation as type IIA orientifold constructions in order to introduce the effective STU-models considered in the rest of the paper\footnote{The STU-models we will discuss correspond to consistent $\textrm{SO}(3)$ truncations of the $\frac{\textrm{SL}(2)\times \textrm{SO}(6,6)}{\textrm{SO}(2) \times \textrm{SO}(6) \times \textrm{SO}(6)}$ coset space spanned by the scalar fields of half-maximal supergravity in four dimensions. The underlying group theory structure guarantees that we are actually solving the full set of equations of motion and not any truncated version thereof, even though we are setting most of the scalars to zero. As usual in supergravity theories (see ref.~\cite{Grana:2013ila} for a recent discussion), the masses of the fields retained in the truncation are not necessarily the lightest ones and therefore the analysis of stability requires the knowledge of the full mass spectrum. We provide the complete spectrum for all vacua discussed in the paper in the appendices.}. In section~\ref{sec:effective actions} we establish the precise correspondence between STU-models and (half-)maximal gauged supergravities in four dimensions. We present the flux/ET dictionary, discuss the interplay between supersymmetry and Scherk-Schwarz conditions as well as the relation to the absence/presence of KK6 monopoles and finally characterise the effective supergravity in terms of the universal moduli powers appearing in the scalar potential. In section \ref{sec:taxonomy} we  exhaustively classify the structure of 4d flux vacua by making a combined use of duality transformations in the STU-models and algebraic geometry techniques in order to solve the extremum conditions of the scalar potential and the consistency relations imposed by supersymmetry. A systematic analysis of the critical points identifying the required sources as well as the underlying $\,\mathcal{N}=4\,$ gauging is performed. We conclude with section~\ref{sec:conclusions} and present some relevant data associated to the M-theory flux vacua in the two appendices.

\section{M-theory on $G_{2}$-manifolds with fluxes}
\label{sec:M-theoryG2}

Our starting point is the Scherk-Schwarz reduction of M-theory on $G_{2}$-manifolds with fluxes derived in ref.~\cite{DallAgata:2005fm}. It is an orbifold reduction on $X_{7}=\frac{\mathbb{T}^7}{\mathbb{Z}_{2} \times \mathbb{Z}_{2} \times \mathbb{Z}_{2}}$ including $G_{(4)}$ and $G_{(7)}$ background fluxes for the $A_{(3)}$ and $A_{(6)}$ gauge potentials of 11d supergravity, as well as a metric $\omega$-flux associated to a twist along the internal space $X_{7}$. We will re-derive the four dimensional effective theory of ref.~\cite{DallAgata:2005fm} in order to establish the set of conventions we are using in this work.

Before introducing the twist, the $G_{2}$-holonomy of the orbifold is encoded in a $G_{2}$ invariant three-form and its 7d dual four-form
\beq
\begin{array}{cclc}
\label{G2_invariant_forms}
\varphi_{G_{2}} &=& dy^{127} + dy^{347} + dy^{567} + dy^{135} - dy^{146} - dy^{362} - dy^{524} & , \\
\star_{7} \, \varphi_{G_{2}} &=& dy^{3456} + dy^{1256} + dy^{1234} - dy^{2467} + dy^{2357} + dy^{4517} + dy^{6137} & ,
\end{array}
\eeq
satisfying  $\,\varphi_{G_{2}} \wedge \star_{7} \, \varphi_{G_{2}}  = 7 \, dy^{1234567}$. We have abbreviated $\,dy^{ABC}\equiv dy^{A} \wedge dy^{B} \wedge dy^{C} \,$ and $\,dy^{ABCD}\equiv dy^{A} \wedge dy^{B} \wedge dy^{C} \wedge dy^{D}\,$ with $A=1,...,7$ in the above expressions. The metric of the internal space is simply the flat metric of the ambient $\mathbb{T}^{7}$, \textit{i.e.} $ds_{7}^2=\sum (\eta^{A})^2$, where $\eta^{A}=R_{A} \, {dy^{A}}$ and $R_{A=1,...,7}$ denote the radii of the seven internal circles. We denote $\,\Phi_{(3)}(R_{A})\,$ the deformed $\varphi_{G_{2}}$ with radii values $R_{A} \neq1$, namely,
\beq
\label{Phi_untwisted}
\Phi_{(3)} = \eta^{127} + \eta^{347} + \eta^{567} + \eta^{135} - \eta^{146} - \eta^{362} - \eta^{524} \ . 
\eeq
Consequently the internal component of the gauge potential $A_{(3)}$ has a similar expansion and both can be combined  into a $G_{2}$ invariant complexified three-form
\beq
\label{A3+Phi_expansion}
\begin{array}{ccll}
\frac{1}{2}  (A_{(3)} + i \Phi_{(3)} ) &=&  \displaystyle\sum_{A=1}^{7} T_{A}(x) \, \omega_{A}(y) & ,
\end{array}
\eeq
where the $\omega_{A}(y)$ entering the above expansion are the seven basis elements of $H^{3}(X_{7})$. The seven coefficients $T_{A}(x)$ represent moduli fields in the four-dimensional effective action.

After a twist is turned on by means of a metric flux\footnote{The $\omega$-metric flux ${\omega_{BC}}^{A}={\omega_{[BC]}}^{A}$ contains the \textbf{140'} (traceless part) and \textbf{7'} irrep's of SL(7) as can be seen from the tensor product $\textbf{21'} \times \textbf{7}  = \textbf{140'} + \textbf{7'}$.}, \textit{i.e.} ${\omega_{BC}}^{A}\neq0$, the $G_{2}$-holonomy of the original orbifold is replaced by a $G_{2}$-structure. The set of left invariant twisted forms $\eta^{A}$ along the internal space then satisfy the Maurer-Cartan equations
\beq
\label{Maurer-Cartan}
d \eta^{A} + \frac{1}{2} {\omega_{BC}}^{A} \eta^{B} \wedge \eta^{C} = 0 \ ,
\eeq
and can be used to build the set $H^{p}(X_{7})$ of cohomology classes of $X_{7}$.

The preserved $G_{2}$-structure ensures $\mathcal{N}=1$ supersymmetry in the reduced theory. The K\"ahler potential for the seven moduli fields $\,T_{A}\,$ in the expansion (\ref{A3+Phi_expansion}) is given by \cite{Beasley:2002db,DallAgata:2005fm}
\beq
\label{Kahler_potential}
K=-\sum_{A=1}^{7} \log\left(  -i ( T_{A}- \bar{T}_{A})\right) \ ,
\eeq
corresponding to a scalar manifold $\,\mathcal{M}_{\textrm{scalar}}=[\textrm{SU}(1,1)/\textrm{U}(1)]^{7}$. In addition a scalar potential also emerges upon reduction -- see  refs~\cite{Gukov:1999gr,Acharya:2000ps,Beasley:2002db} for reductions with $G_{2}$-holonomy and refs~\cite{House:2004pm,DallAgata:2005fm} for weak $G_{2}$-holonomy and cocalibrated \mbox{$G_{2}$-structures} --. This potential can be derived from the flux-induced superpotential \cite{House:2004pm,DallAgata:2005fm}
\beq
\label{WMtheory}
W_{\textrm{M-theory}} = \frac{1}{4} \int_{X_{7}} G_{(7)} + \frac{1}{4} \int_{X_{7}} (A_{(3)} + i \Phi_{(3)}) \wedge \left[ G_{(4)}+\frac{1}{2} d (A_{(3)} + i \Phi_{(3)})\right] \ ,
\eeq
using the standard $\mathcal{N}=1$ supergravity formula
\beq
V = e^{K} \, [ K^{A \bar{B}}  \, D_{A} W \, D_{\bar{B}} \bar{W} - 3 \, W \, \bar{W}   ] \ ,
\eeq
where $K^{A \bar{B}}$ is the inverse of the K\"ahler metric $K_{A \bar{B}}=\partial_{A}\partial_{\bar{B}}K$ and $D_{A} W=\partial_{A} W + (\partial_{A}K) \, W$ is the K\"ahler derivative. The exterior derivative entering the last term in (\ref{WMtheory}) corresponds to the \textit{twisted} operator $d=\partial + \omega$ that incorporates the metric $\omega$-flux in the internal space $X_{7}$. The superpotential (\ref{WMtheory}) consists of three pieces: The first piece is induced by $G_{(7)}$ an produces a constant term. The second piece is induced by $G_{(4)}$ and gives rise to linear couplings for the seven moduli. The third piece is induced by the metric $\omega$-flux and produces quadratic terms $T_{A} T_{B}$ (with $A \neq B$) in the superpotential.

\subsection{M-theory flux-induced superpotential}
\label{section:M-Theory_effective}

Let us now derive the form of the M-theory superpotential (\ref{WMtheory}) in the case of  a reduction on $X_{7}=\mathbb{T}^7/(\mathbb{Z}_{2} \times \mathbb{Z}_{2} \times \mathbb{Z}_{2})$ which has untwisted Betti numbers $b_{1}(X_{7})=b_{2}(X_{7})=0$ and $b_{3}(X_{7})=7$. The geometry of the orbifold is encoded in its sets of invariant forms. Splitting the basis of left invariant twisted 1-forms as
\beq
\eta^{A}=(\,\eta^{a} \, , \, \eta^{i} \, , \, \eta^{7}\,) \ ,
\eeq
with $\,a=1,3,5\,$ and $\,i=2,4,6\,$, then the seven basis elements of $H^{3}(X_{7})$ are given by
\beq
\label{3-form basis}
\begin{array}{llllllll}
\omega_{1}=\eta^{12} \wedge \eta^{7} &\, , \,& \omega_{2}=\eta^{34} \wedge \eta^{7} &\, , \,& \omega_{3}=\eta^{56} \wedge \eta^{7} &\, , \,&  &  \\[2mm]
\alpha_{0}=\eta^{135} &\, , \,& \beta^{1}=\eta^{146}  &\, , \,& \beta^{2}=\eta^{362} &\, , \,& \beta^{3}=\eta^{524} &  .
\end{array}
\eeq
The complementary elements spanning $H^{4}(X_{7})$ are then obtained by 7d Hodge duality and read
\beq
\label{4-form basis}
\begin{array}{llllllll}
\tilde{\omega}^{1}=\eta^{3456} &\, , \,& \tilde{\omega}^{2}=\eta^{1256} &\, , \,& \tilde{\omega}^{3}=\eta^{1234}  &\, , \,& \\[2mm] 
\beta^{0}=\eta^{246} \wedge \eta^{7}  &\, , \,& \alpha_{1}=\eta^{235} \wedge \eta^{7}  &\, , \,& \alpha_{2}=\eta^{451} \wedge \eta^{7} &\, , \,& \alpha_{3}=\eta^{613} \wedge \eta^{7} & .
\end{array}
\eeq
The cohomology basis then satisfies the orthogonality conditions
\beq
\label{orghogonality}
\int_{X_{7}} \omega_{I} \wedge \tilde{\omega}^{J} =  \mathcal{V}_{7} \, \delta_{I}^{J}
\hspace{5mm} , \hspace{5mm}
\int_{X_{7}} \alpha_{0} \wedge \beta^{0} =  - \mathcal{V}_{7} 
\hspace{5mm} , \hspace{5mm}
\int_{X_{7}} \beta^{I} \wedge \alpha_{J} =  -\mathcal{V}_{7} \, \delta^{I}_{J} \ ,
\eeq
with $\,I, J=1,2,3\,$ and where the volume of $X_{7}$ is defined as $\mathcal{V}_{7}=\int_{X_{7}} \eta^{1234567}$.

\begin{table}[t!]
\renewcommand{\arraystretch}{1.35}
\begin{center}
\begin{tabular}{|c|c|c|}
\hline
M-theory origin & Components & Fluxes \\
\hline
\hline
${\omega_{b c}}^{a}$ & $ {\omega_{35}}^{1}\,,\,{\omega_{51}}^{3}\,,\,{\omega_{13}}^{5}$ & 
$\tilde{c}_{1}^{\,(1)}\,,\,\tilde{c}_{1}^{\,(2)}\,,\,\tilde{c}_{1}^{\,(3)}$ \\
\hline
${\omega_{a j}}^{k}$ & $  {\omega_{14}}^{6}\,,\, {\omega_{36}}^{2}\,,\, {\omega_{52}}^{4}$ & 
$\hat{c}_{1}^{\,(1)}\,,\,\hat{c}_{1}^{\,(2)}\,,\,\hat{c}_{1}^{\,(3)}$ \\
\hline
${\omega_{ka}}^{j}$ & $ {\omega_{61}}^{4}\,,\,{\omega_{23}}^{6}\,,\,{\omega_{45}}^{2}$ & 
$\check{c}_{1}^{\,(1)}\,,\,\check{c}_{1}^{\,(2)}\,,\,\check{c}_{1}^{\,(3)}$ \\
\hline
${\omega_{jk}}^{a}$ & ${\omega_{46}}^{1}\,,\,{\omega_{62}}^{3}\,,\,{\omega_{24}}^{5}$ &  
$ b_{1}^{\,(1)}\,,\,b_{1}^{\,(2)}\,,\,b_{1}^{\,(3)}$\\
\hline
$-{\omega_{ai}}^{7}$ & $ -{\omega_{12}}^{7}\,,\,-{\omega_{34}}^{7}\,,\,-{\omega_{56}}^{7}$ & 
$a_{2}^{\,(1)}\,,\,a_{2}^{\,(2)}\,,\,a_{2}^{\,(3)}$ \\
\hline
$-{\omega_{7i}}^{a}$ & $-{\omega_{72}}^{1}\,,\,-{\omega_{74}}^{3}\,,\,-{\omega_{76}}^{5}$ & 
$d_{0}^{\,(1)}\,,\,d_{0}^{\,(2)}\,,\,d_{0}^{\,(3)}$ \\
\hline
$-{\omega_{a7}}^{i}$ & $-{\omega_{17}}^{2}\,,\,-{\omega_{37}}^{4}\,,\,-{\omega_{57}}^{6}$ & 
$c_{3}'^{\,(1)}\,,\,c_{3}'^{\,(2)}\,,\,c_{3}'^{\,(3)}$ \\
\hline
\hline
$-\frac{1}{2} \, G_{aibj}$ & $-\frac{1}{2} \,G_{3456}\,,\,-\frac{1}{2} \,G_{1256}\,,\,-\frac{1}{2} \,G_{1234}$ & 
$a_{1}^{\,(1)}\,,\,a_{1}^{\,(2)}\,,\,a_{1}^{\,(3)}$ \\
\hline
$\frac{1}{2} \,G_{ijk7}$ & $\frac{1}{2} \,G_{2467}$ & 
$b_{0}$ \\
\hline
$\frac{1}{2} \,G_{ibc7}$ & $\frac{1}{2} \,G_{2357}\,,\,\frac{1}{2} \,G_{4517}\,,\,\frac{1}{2} \,G_{6137}$ & 
$c_{0}^{\,(1)}\,,\,c_{0}^{\,(2)}\,,\,c_{0}^{\,(3)}$ \\
\hline
\hline
$\frac{1}{4} \,G_{aibjck7}$ & $\frac{1}{4} \,G_{1234567}$ & 
$a_{0}$ \\
\hline
\end{tabular}
\end{center}
\caption{Metric and gauge fluxes entering the M-theory superpotential.}
\label{Table:M-Theory_fluxes}
\end{table}

Using the above set of invariant forms, it is possible to turn on background fluxes for $G_{(4)}$ and $G_{(7)}$ as well as for the metric $\omega$-flux. In terms of the elements in (\ref{4-form basis}), the $G_{(4)}$ background flux can be expanded as
\beq
\label{G4_fluxes}
\frac{1}{2} \, G_{(4)} = -  \displaystyle\sum_{I} {a_{1}}^{(I)} \, \tilde{\omega}^{I} +  \, b_{0} \, \beta^{0} +  \displaystyle\sum_{I} {c_{0}}^{(I)} \, \alpha_{I} \ .
\eeq
The expansion of the background for $G_{(7)}$ is simply
\beq
\label{G7_fluxes}
\frac{1}{4} \, G_{(7)} =  a_{0} \, \eta^{1234567} \ .
\eeq
In addition to the gauge fluxes (\ref{G4_fluxes}) and (\ref{G7_fluxes}), there are $21$ metric $\omega$-fluxes compatible with the orbifold symmetries. The entire set of M-theory fluxes is summarised in Table~\ref{Table:M-Theory_fluxes}.

In terms of the basis elements (\ref{3-form basis}), the expansion of the complex three-form in (\ref{A3+Phi_expansion}) can be rewritten as
\beq
\label{A3+Phi_expansionSTU}
\begin{array}{ccll}
\frac{1}{2} (A_{(3)} + i \Phi_{(3)}) &=&  \displaystyle\sum_{I} U_{I} \, \omega_{I} + S \, \alpha_{0} - \displaystyle\sum_{I} T_{I} \, \beta^{I}  & ,
\end{array}
\eeq
where $\,S\,$, $\,T_{I}\,$ and $\,U_{I}\,$ have the type IIA interpretation of dilaton, complex structure and K\"ahler moduli, respectively\footnote{Notice the somehow unconventional names for the type IIA moduli fields. We have made this choice in order to exactly reproduce the generalised superpotential of ref.~\cite{Dibitetto:2011gm} derived in the context of type IIB compactifications and further connected to the embedding tensor framework for $\mathcal{N}=4$ supergravity.}. Moreover we also find
\beq
\label{d(A3+Phi)_expansionSTU}
\begin{array}{ccll}
\frac{1}{2}d(A_{(3)} + i \Phi_{(3)})&=& 
\displaystyle\sum_{I} P_{I} \, \tilde{\omega}^{I}  + \beta^{0} \displaystyle\sum_{K} \left( d_{0}^{(K)} T_{K}  - b_{1}^{(K)} U_{K} \right) + \displaystyle\sum_{I} Q^{I} \, \alpha_{I}  \ ,
\end{array}
\eeq
where we have defined the quantities\footnote{In the expressions (\ref{P&Q_quantities}) the $I \neq J \neq K$ assignments have to be understood in a cyclic manner, namely $(I , J , K)=(1 , 2 , 3) \, , \, (2 , 3 , 1) \, , \, (3 , 1 , 2)$. For instance one has $P_{1} = a_{2}^{(2)} U_{3} +a_{2}^{(3)} U_{2} + b_{1}^{(1)} S + \sum_{L}  \mathcal{C}_{1}^{(1L)} \, T_{L}$ and similarly for the rest.}
\beq
\label{P&Q_quantities}
\begin{array}{llll}
P_{I} &=& a_{2}^{(J)} U_{K} +a_{2}^{(K)} U_{J} + b_{1}^{(I)} S + \displaystyle\sum_{L}  \mathcal{C}_{1}^{(IL)} \, T_{L} & \hspace{5mm} (I \neq J \neq K) \ , \\[2mm]
Q^{I} &=& -c_{3}'^{(J)} T_{K} -c_{3}'^{(K)} T_{J} - d_{0}^{(I)} S + \displaystyle\sum_{L}  U_{L} \, \mathcal{C}_{1}^{(LI)} & \hspace{5mm} (I \neq J \neq K) \ ,
\end{array}
\eeq
and where $\mathcal{C}_{1}$ is the flux matrix introduced in ref.~\cite{Font:2008vd}
\beq
\label{C1mat}
\mathcal{C}_{1}^{(IJ)}=\left(
\begin{array}{rrr}
-\tilde{c}_{1}^{\,(1)} & \check{c}_{1}^{\,(3)}   & \hat{c}_{1}^{\,(2)}    \\
 \hat{c}_{1}^{\,(3)}   & -\tilde{c}_{1}^{\,(2)}  & \check{c}_{1}^{\,(1)}  \\
 \check{c}_{1}^{\,(2)} & \hat{c}_{1}^{\,(1)}     & -\tilde{c}_{1}^{\,(3)} \\
\end{array}
\right) \ .
\eeq
By plugging (\ref{G4_fluxes})-(\ref{d(A3+Phi)_expansionSTU}) into the flux-induced superpotential (\ref{WMtheory}) and using the orthogonality conditions (\ref{orghogonality}), one finds the M-theory superpotential
\beq
\label{Superpotential_Flux_MTheory}
\begin{array}{llll}
W_{\textrm{M-theory}} &=& a_{0}  - b_{0} \, S +\displaystyle\sum_{K=1}^{3} c_{0}^{\,(K)} T_{K}  -\displaystyle\sum_{K=1}^{3} a_{1}^{\,(K)}\,U_{K} & \\[2mm]
&+& \displaystyle\sum_{K=1}^{3} a_{2}^{\,(K)} \dfrac{U_{1}U_{2}U_{3}}{U_{K}} + \displaystyle\sum_{I,J=1}^{3}  U_{I} \, \mathcal{C}_{1}^{\,(I J)}\,T_{J} +S \displaystyle\sum_{K=1}^{3} b_{1}^{\,(K)}\,U_{K} \\[2mm]
&-& \displaystyle\sum_{K=1}^{3} c_{3}'^{\,(K)} \dfrac{T_{1}T_{2}T_{3}}{T_{K}} - S \displaystyle\sum_{K=1}^{3} d_{0}^{\,(K)}\,T_{K} & .
\end{array} 
\eeq
With this we conclude the re-derivation of the effective supergravities coming from twisted reductions of M-theory on an $X_{7}=\mathbb{T}^7/(\mathbb{Z}_{2} \times \mathbb{Z}_{2} \times \mathbb{Z}_{2})$ orbifold with fluxes and set up the scenario we will analyse later.

\subsection{$G_{2}$-structure of the M-theory reduction}

The geometry of the twisted $X_{7}=\mathbb{T}^7/(\mathbb{Z}_{2} \times \mathbb{Z}_{2} \times \mathbb{Z}_{2})$ orbifold we are considering determines the set of \mbox{$G_{2}$-structure} relations
\beq
\label{G2_structure}
\begin{array}{ccll}
d \Phi_{(3)}  &=& \widetilde{W}_{1} \,  \star_{7} \Phi_{(3)} + 2 \, \widetilde{W}_{27} & ,  \\[2mm]
d \star_{7} \Phi_{(3)} &=& 0 & ,
\end{array}
\eeq
thus corresponding to a cocalibrated ($\widetilde{W}_{7}=\widetilde{W}_{14}=0$) $G_{2}$-structure \cite{Friedrich:2002,DallAgata:2005fm}. Let us explicitly compute the torsion classes $\,\widetilde{W}_{1}\,$ and $\,\widetilde{W}_{27}$ sitting respectively in the \textbf{1} and \textbf{27} irrep's of G$_{2}$ in the particular case of $A_{(3)}=0$ and $R_{A}=1$ for the seven radii in $X_{7}$. This requires to evaluate the expression (\ref{A3+Phi_expansionSTU}) at the point $\,S=T_{I}=U_{I}=i\,$ so that the twisted versions of the $G_{2}$ invariant forms in (\ref{G2_invariant_forms}) are recovered. These are
\beq
\tfrac{1}{2} \, \Phi_{(3)} =  \displaystyle\sum_{I=1}^{3}  \omega_{I} + \alpha_{0} - \displaystyle\sum_{I=1}^{3} \beta^{I}
\hspace{5mm} , \hspace{5mm} 
\tfrac{1}{2} \, \star_{7}  \, \Phi_{(3)} = \displaystyle\sum_{I=1}^{3}  \tilde{\omega}_{I} -  \beta^{0} + \displaystyle\sum_{I=1}^{3} \alpha_{I} \ ,
\eeq
with the non-standard normalisation $\,\tfrac{1}{7} \int_{X_{7}} \Phi_{(3)} \wedge \star_{7} \Phi_{(3)} =  4 \, \mathcal{V}_{7}\,$. After some algebra we obtain a  one-parameter family -- the non-trivial condition in (\ref{G2_structure}) is linear -- of torsion classes satisfying (\ref{G2_structure}). It is given by
\beq
\label{G2_torsion_family}
\widetilde{W}_{1}=(1+\kappa) \, W_{1}
\hspace{10mm} \textrm{ and } \hspace{10mm}
2\, \widetilde{W}_{27}= 2 \, W_{27} - \kappa \, W_{1} \, \star_{7} \Phi_{(3)} \ ,
\eeq
where we have introduced the flux-dependent quantities
\beq
\label{G2_torsion}
\begin{array}{ccll}
W_{1} & = & \displaystyle \sum_{L} a_{2}^{(L)} + \displaystyle \sum_{L} b_{1}^{(L)} + \displaystyle \sum_{IJ} \mathcal{C}_{1}^{(IJ)} - \displaystyle \sum_{L} d_{0}^{(L)} - \displaystyle \sum_{L} c_{3}'^{(L)} & , \\[3mm]
W_{27} & = & \displaystyle \sum_{I} A_{I} \, \tilde{\omega}^{I} + B \, \beta^{0}  + \displaystyle \sum_{I} C_{I} \, \alpha_{I} & .
\end{array}
\eeq
The coefficients in the expansion of $W_{27}$ also depend on the flux parameters and read\footnote{As in (\ref{P&Q_quantities}), the $I \neq J \neq K$ assignments are understood in a cyclic manner also in (\ref{W27_torsion_components}). This time one has $A_{1}  =  -a_{2}^{(1)} - b_{1}^{(2)} - b_{1}^{(3)} - \sum_{L} ( \mathcal{C}_{1}^{(2L)} +  \mathcal{C}_{1}^{(3L)} ) + \sum_{L} ( d_{0}^{(L)} +  c_{3}'^{(L)} )$ and similarly for the others.}
\beq
\label{W27_torsion_components}
\begin{array}{ccll}
A_{I} & = & -a_{2}^{(I)} - b_{1}^{(J)} - b_{1}^{(K)} - \displaystyle \sum_{L} ( \mathcal{C}_{1}^{(JL)} +  \mathcal{C}_{1}^{(KL)} ) + \displaystyle \sum_{L} ( d_{0}^{(L)} +  c_{3}'^{(L)} ) &  (I \neq J \neq K)  \\[3mm] 
B & = & \displaystyle \sum_{L} a_{2}^{(L)}  + \displaystyle \sum_{IJ}  \mathcal{C}_{1}^{(IJ)}  - \displaystyle \sum_{L} c_{3}'^{(L)}   &  \\[3mm]
C_{I} & = & c_{3}'^{(I)} + d_{0}^{(J)} + d_{0}^{(K)} - \displaystyle \sum_{L} ( \mathcal{C}_{1}^{(LJ)} +  \mathcal{C}_{1}^{(LK)} ) - \displaystyle \sum_{L} ( a_{2}^{(L)} +  b_{1}^{(L)} ) &  (I \neq J \neq K)  
\end{array}
\eeq
In order to recover the standard $G_{2}$ relations for the properly normalised $\Phi_{(3)}$ and $\star_{7} \Phi_{(3)}$ forms \cite{Bryant:2003,Held:2011uz}
\beq
\label{Torsion_standard}
\widetilde{W}_{1}=\frac{1}{7} \, (\tfrac{1}{2} \, d\Phi_{(3)}) \lrcorner (\tfrac{1}{2} \, \star_{7} \Phi_{(3)})  \hspace{5mm} \textrm{ and } \hspace{5mm} \widetilde{W}_{27}=(\tfrac{1}{2} \, d\Phi_{(3)}) - \widetilde{W}_{1} \,   (\tfrac{1}{2} \, \star_{7} \Phi_{(3)})\ , 
\eeq
one must set the parameter $\kappa=-5/7$ in (\ref{G2_torsion_family}). Up to an overall $\frac{1}{16}$ factor coming from the normalisation of (\ref{WMtheory}), this is consistent with the relation \cite{Held:2011uz} between the potential energy induced by the metric $\omega$-flux and the Ricci scalar of $X_{7}$ 
\beq
V_{\omega} = -\frac{1}{16} \, \textrm{Ricci}_{X_{7}} = -\frac{1}{16} \, \left( \frac{21}{8} \, |\widetilde{W}_{1}|^2 - \frac{1}{2} \, |\widetilde{W}_{27}|^2 \right) \ .
\eeq

Generic M-theory flux vacua will activate the two torsion classes $\,\widetilde{W}_{1}\,$ and $\,\widetilde{W}_{27}\,$ thus specifying a cocalibrated $G_{2}$-structure. However under certain circumstances -- for instance at ${\mathcal{N}=1}$ supersymmetric AdS$_4$ solutions \cite{DallAgata:2005fm} -- one might have  $\,\widetilde{W}_{27}=0\,$ determining a weak \mbox{$G_{2}$-holonomy} or even $\,\widetilde{W}_{1}=\widetilde{W}_{27}=0\,$ restoring a $G_{2}$-holonomy. We will investigate this issue for the set of M-theory flux vacua we will obtain in section~\ref{sec:taxonomy}.

\subsection{Interpretation as type IIA orientifolds}

Massless type IIA supergravity can be obtained from reduction of M-theory along the 11th direction. Schematically, this amounts to the splitting
\beq
X_{7} = \frac{\mathbb{T}^{7}}{\mathbb{Z}_{2} \times \mathbb{Z}_{2} \times \mathbb{Z}_{2}}  \,\, \longrightarrow \,\, 
\frac{\mathbb{T}^{6}}{\mathbb{Z}_{2} \times \mathbb{Z}_{2}} \times \eta^{7} = X_{6} \times \eta^{7} \ ,
\eeq
with $\eta^{m=1,...,6}$ being associated to $X_{6}\,$, additionally endowed with an extra $\mathbb{Z}_{2}$ ``orientifold" involution reflecting the coordinates $\eta^{i} \rightarrow -\eta^{i}$ and $\eta^{7} \rightarrow -\eta^{7}$. This is compatible with the following invariant forms. From the forms in (\ref{3-form basis}) and (\ref{4-form basis}), one reads off the elements spanning $H^{2}(X_{6})$
\beq
\label{2-form basisIIA}
\omega_{1}=\eta^{12}  \,\,\,\,\,\,\, , \,\,\,\,\,\,\, \omega_{2}=\eta^{34}  \,\,\,\,\,\,\, , \,\,\,\,\,\,\, \omega_{3}=\eta^{56} \ ,
\eeq
as well as those spanning $H^{4}(X_{6})$
\beq
\label{4-form basisIIA}
\tilde{\omega}^{1}=\eta^{3456}  \,\,\,\,\,\,\, , \,\,\,\,\,\,\, \tilde{\omega}^{2}=\eta^{1256}  \,\,\,\,\,\,\, , \,\,\,\,\,\,\, \tilde{\omega}^{3}=\eta^{1234} \ .
\eeq
Similarly, the elements spanning $H^{3}(X_{6})$ are given by
\beq
\label{3-form basisIIA}
\begin{array}{llllllll}
\alpha_{0}=\eta^{135} &\, , \,& \beta^{1}=\eta^{146}  &\, , \,& \beta^{2}=\eta^{362} &\, , \,& \beta^{3}=\eta^{524} & , \\[2mm]
\beta^{0}=\eta^{246}  &\, , \,& \alpha_{1}=\eta^{235}  &\, , \,& \alpha_{2}=\eta^{451}  &\, , \,& \alpha_{3}=\eta^{613}  & .
\end{array}
\eeq
The volume of $X_{6}$ is then defined as $\mathcal{V}_{6}=\int_{X_{6}} \eta^{123456}$ and the orthogonality conditions in (\ref{orghogonality}) tranlates into
\beq
\label{orghogonality_IIA}
\int_{X_{6}} \omega_{I} \wedge \tilde{\omega}^{J} =  \mathcal{V}_{6} \, \delta_{I}^{J}
\hspace{5mm} , \hspace{5mm}
\int_{X_{6}} \alpha_{0} \wedge \beta^{0} =  - \mathcal{V}_{6} 
\hspace{5mm} , \hspace{5mm}
\int_{X_{6}} \beta^{I} \wedge \alpha_{J} =  -\mathcal{V}_{6} \, \delta^{I}_{J} \ .
\eeq

Analogously to the M-theory case, background fluxes for \textit{all} the type IIA gauge potentials can be turned on together with a type IIA metric $\omega$-flux\footnote{The type IIA metric $\omega$-flux ${\omega_{np}}^{m}={\omega_{[np]}}^{m}$ contains the \textbf{84'} (traceless part) and \textbf{6'} irrep's of SL(6) as can be seen from the tensor product $\textbf{15'} \times \textbf{6}  = \textbf{84'} + \textbf{6'}$. They descend from the original M-theory $\omega$-fluxes by virtue of the $\textrm{SL}(7) \supset \textrm{SL}(6)$ decompositions $\textbf{140'} \rightarrow \textbf{84'} + \textbf{15'} + \textbf{35} + \textbf{6'}$ and  $\textbf{7'} \rightarrow  \textbf{6'} + \textbf{1}$ which can be equivalently viewed as $\,{\omega_{BC}}^{A} \rightarrow {\omega_{np}}^{m} \oplus {\omega_{np}}^{7} \oplus {\omega_{7p}}^{m} \oplus {\omega_{7p}}^{7}\,$ and $\,{\omega_{BC}}^{C} \rightarrow {\omega_{nC}}^{C} \oplus {\omega_{7C}}^{C}$. The orbifold symmetries reduce the number of IIA metric fluxes $\,{\omega_{np}}^{m}\,$ to $12$.} in the internal space $X_{6}$. In terms of the cohomology basis (\ref{2-form basisIIA})-(\ref{3-form basisIIA}), the R-R background fluxes can be expanded as
\beq
\label{RR-fluxes}
F_{(6)} = a_{0} \, \eta^{123456}
\hspace{4mm} , \hspace{4mm}
F_{(4)} = -a_{1}^{(I)} \, \tilde{\omega}^{I}
\hspace{4mm} , \hspace{4mm}
F_{(2)} = a_{2}^{(I)} \, \omega_{I}
\hspace{4mm} , \hspace{4mm}
F_{(0)} = -a_{3}
\eeq
whereas the expansion of the NS-NS flux can be taken as
\beq
\label{NSNS-fluxes}
H_{(3)} = b_{0} \, \beta^{0}  + c_{0}^{(I)} \, \alpha_{I} \ . 
\eeq
Importantly, the $F_{(0)}=-a_{3}$ flux parameter in (\ref{RR-fluxes}) corresponds to the Romans mass in \textit{massive} type IIA supergravity \cite{Romans:1985tz} and \textit{does not} directly descend from M-theory. The full set of type IIA fluxes including also metric fluxes is summarised in Table~\ref{Table:IIA_fluxes}.

\begin{table}[t!]
\renewcommand{\arraystretch}{1.35}
\begin{center}
\begin{tabular}{|c|c|c|}
\hline
Type IIA origin & Components & Fluxes \\
\hline
\hline
${\omega_{b c}}^{a}$ & $ {\omega_{35}}^{1}\,,\,{\omega_{51}}^{3}\,,\,{\omega_{13}}^{5}$ & 
$\tilde{c}_{1}^{\,(1)}\,,\,\tilde{c}_{1}^{\,(2)}\,,\,\tilde{c}_{1}^{\,(3)}$ \\
\hline
${\omega_{a j}}^{k}$ & $  {\omega_{14}}^{6}\,,\, {\omega_{36}}^{2}\,,\, {\omega_{52}}^{4}$ & 
$\hat{c}_{1}^{\,(1)}\,,\,\hat{c}_{1}^{\,(2)}\,,\,\hat{c}_{1}^{\,(3)}$ \\
\hline
${\omega_{ka}}^{j}$ & $ {\omega_{61}}^{4}\,,\,{\omega_{23}}^{6}\,,\,{\omega_{45}}^{2}$ & 
$\check{c}_{1}^{\,(1)}\,,\,\check{c}_{1}^{\,(2)}\,,\,\check{c}_{1}^{\,(3)}$ \\
\hline
${\omega_{jk}}^{a}$ & ${\omega_{46}}^{1}\,,\,{\omega_{62}}^{3}\,,\,{\omega_{24}}^{5}$ &  
$ b_{1}^{\,(1)}\,,\,b_{1}^{\,(2)}\,,\,b_{1}^{\,(3)}$\\
\hline
${F_{ai}}$ & $ F_{12}\,,\,F_{34}\,,\,F_{56}$ & 
$a_{2}^{\,(1)}\,,\,a_{2}^{\,(2)}\,,\,a_{2}^{\,(3)}$ \\
\hline
\multicolumn{2}{|c|}{non-geometric}  & 
$d_{0}^{\,(1)}\,,\,d_{0}^{\,(2)}\,,\,d_{0}^{\,(3)}$ \\
\hline
\multicolumn{2}{|c|}{non-geometric} & 
$c_{3}'^{\,(1)}\,,\,c_{3}'^{\,(2)}\,,\,c_{3}'^{\,(3)}$ \\
\hline
\hline
$-F_{aibj}$ & $-F_{3456}\,,\,-F_{1256}\,,\,-F_{1234}$ & 
$a_{1}^{\,(1)}\,,\,a_{1}^{\,(2)}\,,\,a_{1}^{\,(3)}$ \\
\hline
$H_{ijk}$ & $H_{246}$ & 
$b_{0}$ \\
\hline
$H_{ibc}$ & $H_{235}\,,\,H_{451}\,,\,H_{613}$ & 
$c_{0}^{\,(1)}\,,\,c_{0}^{\,(2)}\,,\,c_{0}^{\,(3)}$ \\
\hline
\hline
$F_{aibjck}$ & $F_{123456}$ & 
$a_{0}$ \\
\hline
\hline
\multicolumn{2}{|c|}{$-F_{(0)}$ \hspace{2mm} (Romans mass)} & 
$a_{3}$ \\
\hline
\end{tabular}
\end{center}
\caption{Metric, gauge and non-geometric fluxes entering the type IIA superpotential.}
\label{Table:IIA_fluxes}
\end{table}

In type IIA orientifold compactifications including O6-planes and D6-branes, the flux-induced superpotential takes the form \cite{Derendinger:2004jn,Villadoro:2005cu}
\beq
\label{WIIA}
W_{\textrm{IIA}} = \int_{X_{6}} e^{J_{c}} \wedge F + \int_{X_{6}} \Omega_{c} \wedge (H_{(3)} + dJ_{c}) \ ,
\eeq
where the (complexified) K\"ahler two-form $J_{c}$ and the holomorphic three-form $\Omega_{c}$ can be read off from (\ref{A3+Phi_expansionSTU}) by requiring $\,\frac{1}{2} (A_{(3)} + i \Phi_{(3)}) = J_{c} \wedge \eta^{7} + \Omega_{c}\,$. This is
\beq
\label{Jc_OmegaSTU}
J_{c} =  \displaystyle\sum_{I} U_{I} \, \omega_{I}
\hspace{10mm} \textrm{ and } \hspace{10mm} 
\Omega_{c} = S \, \alpha_{0} \, - \,  \displaystyle\sum_{I} T_{I} \, \beta^{I} \ .
\eeq
Using the type IIA metric $\omega$-fluxes in $X_{6}$ displayed in Table~\ref{Table:IIA_fluxes} one finds
\beq
\label{dJc_expansionSTU}
\begin{array}{ccll}
dJ_{c}&=&  - \beta^{0} \, \displaystyle\sum_{K} b_{1}^{(K)} U_{K}  + \alpha_{I} \, \displaystyle\sum_{L}  U_{L} \, \mathcal{C}_{1}^{(LI)}   \ ,
\end{array}
\eeq
and an explicit computation of the superpotential (\ref{WIIA}) yields
\beq
\label{Superpotential_FluxIIA}
\begin{array}{llll}
W_{\textrm{IIA}} &=& a_{0}  - b_{0} \, S + \displaystyle\sum_{K=1}^{3} c_{0}^{\,(K)} T_{K}  -\displaystyle\sum_{K=1}^{3} a_{1}^{\,(K)}\,U_{K} & \\[2mm]
&+& \displaystyle\sum_{K=1}^{3} a_{2}^{\,(K)} \dfrac{U_{1}U_{2}U_{3}}{U_{K}} + \displaystyle\sum_{I,J=1}^{3}  U_{I} \, \mathcal{C}_{1}^{\,(I J)}\,T_{J} +S \displaystyle\sum_{K=1}^{3} b_{1}^{\,(K)}\,U_{K} -a_{3} U_{1} U_{2} U_{3} \ .
\end{array} 
\eeq

As noticed in refs~\cite{DallAgata:2005fm,Aldazabal:2006up}, the ordinary type IIA orientifold reductions including gauge plus metric fluxes miss the $\,c_{3}'^{(I)}\,$ and $\,d_{0}^{(I)}\,$ fluxes with respect to the ordinary M-theory construction of the previous sections. However, they gain the Romans mass parameter $a_{3}$ and the corresponding cubic coupling in the IIA superpotential (\ref{Superpotential_FluxIIA}). As a consequence the M-theory superpotential (\ref{Superpotential_Flux_MTheory}) can be viewed as a \textit{massless} ($a_{3}=0$) but \textit{generalised}  type IIA superpotential including the \textit{non-geometric} fluxes $c_{3}'^{(I)}$ and $d_{0}^{(I)}$ which induce the last two terms in (\ref{Superpotential_Flux_MTheory}). The situation can be described as follows
\beq
\label{Superpotential_comparison}
W_{\textrm{M-theory}} \,\,=\,\, W^{(a_{3}=0)}_{\textrm{IIA}} + W_{\textrm{non-geom}} \,\,=\,\, W^{(a_{3}=0)}_{\textrm{IIA}} - \displaystyle\sum_{K=1}^{3} c_{3}'^{\,(K)} \dfrac{T_{1}T_{2}T_{3}}{T_{K}} - S \displaystyle\sum_{K=1}^{3} d_{0}^{\,(K)}\,T_{K} \ .
\eeq
We will elaborate more on the consequences of turning on these type IIA \textit{non-geometric} fluxes $c_{3}'^{(I)}$ and $d_{0}^{(I)}$ as well as on the interpretation of the corresponding flux-induced vacua as backgrounds containing KK monopoles, thus going beyond twisted tori as suggested in ref.~\cite{DallAgata:2005fm} (see discussion in section~$5.2$ therein). Our approach here will be completely four-dimensional as we will be using the effective theory of $\mathcal{N}=4$ gauged supergravity \cite{Schon:2006kz} as the theoretical framework in which to describe the backgrounds.

\subsection{Cyclic symmetry and STU-models}

In order to simplify the setup as much as possible we will further restrict to the \textit{isotropic} scenario in which a cyclic SO(3) symmetry $I \rightarrow J \rightarrow K $ is imposed \cite{Derendinger:2004jn}. This simplification is compatible with an Ansatz
\beq
\label{LimIsoFields}
T_{1}=T_{2}=T_{3} \equiv T \hspace{10mm} \textrm{ and } \hspace{10mm}U_{1}=U_{2}=U_{3} \equiv U
\eeq
for the four-dimensional moduli fields. The K\"ahler potential in (\ref{Kahler_potential}) then reduces to the isotropic form
\beq
\label{Kahler_potential_ISO}
K^{(\textrm{iso})}=-\log\left(  -i ( S- \bar{S})\right) -3 \, \log\left(  -i ( T- \bar{T})\right)  - 3 \, \log\left(  -i ( U- \bar{U})\right)  \ ,
\eeq
which corresponds to a $\,\mathcal{M}_{\textrm{scalar}}=[\textrm{SU}(1,1)/\textrm{U}(1)]_{S} \times [\textrm{SU}(1,1)/\textrm{U}(1)]_{T}\times[\textrm{SU}(1,1)/\textrm{U}(1)]_{U}\,$  manifold described by the moduli fields of the so-called STU-model. This simplification is also consistent with an \textit{isotropic} flux Ansatz of the form
\beq
\label{LimIsoFluxes_1}
\tilde{c}_{1}^{(I)}=\tilde{c}_{1} \,\,\,\, , \,\,\,\, \hat{c}_{1}^{(I)}=\check{c}_{1}^{(I)}=c_{1} \,\,\,\, , \,\,\,\, b_{1}^{(I)}=b_{1} \,\,\,\, , \,\,\,\, a_{2}^{(I)}=a_{2} \,\,\,\, , \,\,\,\, d_{0}^{(I)}=d_{0} \,\,\,\, , \,\,\,\, c_{3}'^{(I)}=c_{3}'  
\eeq
for the M-theory metric $\omega$-fluxes in Table~\ref{Table:M-Theory_fluxes} and similarly for the gauge fluxes
\beq
\label{LimIsoFluxes_2}
a_{1}^{(I)}=a_{1} \,\,\,\,\,\,\, , \,\,\,\,\,\,\, c_{0}^{(I)}=c_{0} \ .
\eeq
The above content of fields and fluxes has been shown to be part of the SO(3) invariant sector of the maximal and half-maximal supergravities in four dimensions, the latter being coupled to six vector multiplets \cite{Dibitetto:2011gm,Dibitetto:2012ia}. We will exploit this fact later on in the paper to investigate the effect of introducing M-theory monopoles in the compactification scheme.

In the isotropic limit, the expression (\ref{Superpotential_Flux_MTheory}) of the M-theory flux-induced superpotential takes the form
\beq
\label{Superpotential_Flux_MTheory_ISO}
\begin{array}{llll}
W^{\textrm{(iso)}}_{\textrm{M-theory}} &=& a_{0}  - b_{0} \, S + 3 \, c_{0} \, T  - 3 \, a_{1} \,U + 3 \, a_{2} \, U^{2} + 3 \,(2 \, c_{1} - \tilde{c}_{1}) \, U\,T  + 3 \,b_{1} \,  S \, U \\[2mm]
&-& 3 \, c_{3}' \, T^{2} - 3 \, d_{0} \,  S \, T  \ ,
\end{array} 
\eeq
whereas the type IIA superpotential in (\ref{Superpotential_FluxIIA}) reduces to \cite{Derendinger:2004jn}
\beq
\label{Superpotential_Flux_IIA_ISO}
\begin{array}{rlll}
\,\,\,\,\,\,\,W^{\textrm{(iso)}}_{\textrm{IIA}} &=& a_{0}  - b_{0} \, S + 3 \, c_{0} \, T  - 3 \, a_{1} \,U + 3 \, a_{2} \, U^{2} + 3 \,(2 \, c_{1} - \tilde{c}_{1}) \, U\,T  + 3 \,b_{1} \,  S \, U \\[2mm]
&-& a_{3} \, U^{3} \ .
\end{array}
\eeq
%
These are the M-theory and type IIA superpotentials we will consider during the rest of the paper. Notice that the relation (\ref{Superpotential_comparison}) still holds in its isotropic version
\beq
\label{Superpotential_comparison_ISO}
W_{\textrm{M-theory}} ^{\textrm{(iso)}}\,\,=\,\, W^{\textrm{(iso)} (a_{3}=0)}_{\textrm{IIA}} + W^{\textrm{(iso)}}_{\textrm{non-geom}} \,\,=\,\, W^{\textrm{(iso)} (a_{3}=0)}_{\textrm{IIA}} - 3 \, c_{3}'\, T^{2} - 3 \, d_{0} \, S  \,T  \ ,
\eeq
making the connection between M-theory and type IIA effective STU-models manifest.

The simplifications (\ref{LimIsoFluxes_1}) and (\ref{LimIsoFluxes_2}) on the fluxes also translate into simpler torsion classes $\widetilde{W}_{1}$ and $\,\widetilde{W}_{27}\,$ specifying the \textit{isotropic}  $\,G_{2}$-structure. The expressions (\ref{G2_torsion}) and (\ref{W27_torsion_components}) simplify to
\beq
\label{G2_torsion_ISO}
\begin{array}{ccll}
W_{1} & = & 3\,  a_{2} + 3\,  b_{1} + 3 \, (2 c_{1} - \tilde{c}_{1})  - 3\,  d_{0}  - 3 \, c_{3}'  & , \\[1mm]
W_{27} & = & A \displaystyle \sum_{I}  \, \tilde{\omega}^{I} + B \, \beta^{0}  + C \displaystyle \sum_{I} \, \alpha_{I} & ,
\end{array}
\eeq
with the flux-dependent coefficients in $W_{27}$ given by
\beq
\label{W27_torsion_components_ISO}
\begin{array}{lcll}
A & = & -a_{2} - 2 \, b_{1} -  2\, (2 c_{1} - \tilde{c}_{1}) + 3\,  d_{0} + 3\, c_{3}'  & , \\ 
B & = & 3\,  a_{2}  + 3 \, (2 c_{1} - \tilde{c}_{1})  - 3 \,  c_{3}'   &  , \\
C & = & c_{3}' + 2 \, d_{0}  - 2\, (2 c_{1} - \tilde{c}_{1}) -  3 \, a_{2} - 3 \,  b_{1} & .
\end{array}
\eeq
Constraining the torsion classes, \textit{e.g.} demanding $\,\widetilde{W}_{27}=0\,$ to have weak $G_{2}$-holonomy, imposes linear relations on the background fluxes that simplify the resulting STU-models.

\section{Effective action and gauged supergravitites}
\label{sec:effective actions}

In this section we investigate the connection between the consistency conditions in Scherk-Schwarz reductions of M-theory (top-down) and the consistency conditions in effective $\,{\mathcal{N}=4}\,$ and  $\,\mathcal{N}=8\,$ gauged supergravities (bottom-up). We will link such conditions to the absence/presence of KK6 monopoles in the M-theory background and characterise the resulting scalar potential in the effective supergravity action.

\subsection{Scherk-Schwarz reductions and BI}

The M-theory fluxes are restricted by a set of quadratic constraints coming from the consistency of the reduction down to four dimensions  \cite{Scherk:1979zr,Dall'Agata:2005ff,Dall'Agata:2005mj,DallAgata:2005fm}. In an ordinary Scherk-Schwarz reduction of M-theory these are
\beq
\label{SS_Constraints}
{\omega_{[AB}}^{F} \, {\omega_{C]F}}^{D}  = 0
\hspace{10mm} \textrm{ and } \hspace{10mm}
{\omega_{[AB}}^{F} \, G_{CDE]F}  = 0
\eeq
coming respectively from the nilpotency ($d^2=0$) of the \textit{twisted} derivative operator ${d=\partial + \omega}$ as well as from the \textit{twisted} Bianchi identity (BI) $\,dG_{(4)}=0\,$ along the internal space $X_{7}$. Moreover the symmetries of the $X_{7}=\mathbb{T}^7/(\mathbb{Z}_{2} \times \mathbb{Z}_{2} \times \mathbb{Z}_{2})$ orbifold guarantees $\,{\omega_{AB}}^{A}=0$ (compact $X_{7}$ with no boundary), thus implying a well-defined Lagrangian upon reduction \cite{Scherk:1979zr}.

The first quadratic constraint in (\ref{SS_Constraints}) gives rise to a set of $\,6+6+3+1+3+6+3=28\,$ conditions of the form\footnote{In the expressions (\ref{ww_constraint}) the $I \neq J \neq K$ assignments are understood in two different manners. For conditions coming in a triplet (multiplicity 3) they are understood in a cyclic manner as before, namely ${(I , J , K)=(1 , 2 , 3) \, , \, (2 , 3 , 1) \, , \, (3 , 1 , 2)}$. For conditions coming in a sextuplet (multiplicity 6) they are understood as permutations, namely $(I , J , K)=(1 , 2 , 3) \, , \, (2 , 1 , 3) \, , \,  (2 , 3 , 1) \, , \,  (3 , 2 , 1) \, , \, (3 , 1 , 2) \, , \, (1 , 3 , 2)$.}
\beq
\label{ww_constraint}
\begin{array}{rrlrllllll}
i) & {\omega_{[ai}}^{D} \, {\omega_{c]D}}^{k}  = 0 &  \rightarrow &   -a_{2}^{(I)} c_{3}'^{(J)} + \mathcal{C}_{1}^{(KK)}  \mathcal{C}_{1}^{(JI)}   + \mathcal{C}_{1}^{(JK)} \mathcal{C}_{1}^{(KI)} =  0 & \,\,\, (I\neq J \neq K)  \\[2mm]
ii) & {\omega_{[ai}}^{D} \, {\omega_{k]D}}^{c}  = 0 &  \rightarrow &   -d_{0}^{(I)} a_{2}^{(J)} + \mathcal{C}_{1}^{(II)}  b_{1}^{(K)}   + \mathcal{C}_{1}^{(KI)} b_{1}^{(I)} =  0 & \,\,\, (I\neq J \neq K)  \\[2mm]
iii)  & {\omega_{[ib}}^{D} \, {\omega_{c]D}}^{7}  = 0 &  \rightarrow &   \displaystyle \sum_{L} a_{2}^{(L)} \mathcal{C}_{1}^{(LI)}  =  0 &   \\[2mm]
iv)  & {\omega_{[ij}}^{D} \, {\omega_{k]D}}^{7}  = 0 &  \rightarrow &   \displaystyle \sum_{K} b_{1}^{(K)} a_{2}^{(K)}    =  0 &   \\[2mm]
v) & {\omega_{[7a}}^{D} \, {\omega_{b]D}}^{k}  = 0 &  \rightarrow &   \displaystyle \sum_{L}  \mathcal{C}_{1}^{(IL)} c_{3}'^{(L)}  =  0 &   \\[2mm]
vi)  & {\omega_{[7a}}^{D} \, {\omega_{j]D}}^{c}  = 0 &  \rightarrow &   b_{1}^{(I)} c_{3}'^{(J)} + \mathcal{C}_{1}^{(II)}  d_{0}^{(K)}   + \mathcal{C}_{1}^{(IK)} d_{0}^{(I)} =  0 & \,\,\, (I\neq J \neq K) \\[2mm]

vii) & {\omega_{[7i}}^{D} \, {\omega_{j]D}}^{k}  = 0 &  \rightarrow &   b_{1}^{(I)} c_{3}'^{(I)} + \mathcal{C}_{1}^{(IJ)}  d_{0}^{(K)}   + \mathcal{C}_{1}^{(IK)} d_{0}^{(J)} =  0 & \,\,\, (I\neq J \neq K)  
\end{array}
\eeq
whereas the second quadratic constraint in (\ref{SS_Constraints}) is automatically satisfied due to the orbifold symmetries. This can be straightforwardly verified using the M-theory fluxes in Table~\ref{Table:M-Theory_fluxes}.

The application of the isotropic limits (\ref{LimIsoFluxes_1}) and (\ref{LimIsoFluxes_2}) to the flux parameters reduces the set of quadratic constraints in (\ref{ww_constraint}) to only $7$ conditions. These are given by
\beq
\label{ww_constraint_iso}
\begin{array}{rrlrllllll}
i) & {\omega_{[ai}}^{D} \, {\omega_{c]D}}^{k}  = 0 &  \rightarrow &   - a_{2} \, c_{3}' + c_{1}\,  (c_{1}-\tilde{c}_{1}) =  0 &   \\[2mm]
ii) & {\omega_{[ai}}^{D} \, {\omega_{k]D}}^{c}  = 0 &  \rightarrow &   -d_{0} \, a_{2} + (c_{1}-\tilde{c}_{1}) \, b_{1}  =  0 &   \\[2mm]
iii)  & {\omega_{[ib}}^{D} \, {\omega_{c]D}}^{7}  = 0 &  \rightarrow & a_{2} \,  (2 \, c_{1}-\tilde{c}_{1})  =  0 &   \\[2mm]
iv)  & {\omega_{[ij}}^{D} \, {\omega_{k]D}}^{7}  = 0 &  \rightarrow &   3\,  b_{1}\, a_{2}    =  0 &   \\[2mm]
v) & {\omega_{[7a}}^{D} \, {\omega_{b]D}}^{k}  = 0 &  \rightarrow &  (2 \, c_{1}-\tilde{c}_{1})  \,  c_{3}'  =  0 &   \\[2mm]
vi)  & {\omega_{[7a}}^{D} \, {\omega_{j]D}}^{c}  = 0 &  \rightarrow &   b_{1} c_{3}' +  (c_{1}-\tilde{c}_{1})  \, d_{0} =  0 & \\[2mm]
vii) & {\omega_{[7i}}^{D} \, {\omega_{j]D}}^{k}  = 0 &  \rightarrow &   b_{1} c_{3}' + 2 \, c_{1} \,  d_{0} =  0 &   .
\end{array}
\eeq
We will investigate the connection between the set of quadratic constraints in (\ref{ww_constraint_iso}) and those required if demanding $\,\mathcal{N}=8\,$ or $\,\mathcal{N}=4\,$ supersymmetry in the effective action. We will discuss it in the framework of the embedding tensor \cite{Schon:2006kz}.

\subsection{Extended supersymmetry and gaugings}

The M-theory superpotential in (\ref{Superpotential_Flux_MTheory}) is an holomorphic function of the moduli fields and therefore completely unrestricted from the point of view of $\mathcal{N}=1$ supergravity. However, a higher-dimensional origin as an ordinary Scherk-Schwarz reduction of M-theory demands the additional constraints in (\ref{ww_constraint}) to be satisfied. We will show now that these conditions are in one-to-one correspondence with the quadratic constraints on the embedding tensor of $\mathcal{N}=8$ supergravity.

Let us start with an intermediate theory between minimal $\,\mathcal{N}=1\,$ and maximal $\,\mathcal{N}=8\,$ supergravity: the half-maximal $\mathcal{N}=4$ supergravity theory coupled to six vector multiplets. This theory has a global symmetry group $\,G=\textrm{SL}(2) \times \textrm{SO}(6,6)\,$ reflecting the putative S and T dualities of string theory upon toroidal reduction. From a purely supergravity point of view, the flux parameters entering the M-theory superpotential (\ref{Superpotential_Flux_MTheory}) determine what is called a \textit{gauging} or deformation of the $\,\mathcal{N}=4\,$ free theory. After applying a gauging, a non-abelian gauge symmetry $\,G_{0}\subset G\,$ emerges in the effective action. The gauge algebra is specified by the commutation relations
\beq
\label{BracketsN=4}
\left[ T_{\alpha M} \, T_{\beta N} \right]  = {f_{\alpha MN}}^{P} \, T_{\beta P} \ ,
\eeq
where $\,T_{\alpha M}\,$ denotes the generators associated to the non-abelian vector fields -- indices $\,\alpha=+,-\,$ and $\,M=1,...,12\,$ are respectively fundamental SL(2) and SO(6,6) indices -- and $\,{f_{\alpha MN}}^{P}\,$ (structure constants) is the so-called \textit{embedding tensor} (ET).

The \mbox{M-theory} fluxes in (\ref{Superpotential_Flux_MTheory}) can be mapped to different components of the embedding tensor. To be more precise, this connection was established \cite{Dall'Agata:2009gv,Dibitetto:2011gm} in a type IIA (and also IIB) incarnation of the four-dimensional STU-model defined by (\ref{Superpotential_Flux_MTheory}). Using light-cone coordinates for the SO(6,6) fundamental index $\,M\,$ amounts to choosing
\beq
\eta = 
\left(
\begin{array}{cc}
0 & \mathbb{I}_{6} \\
\mathbb{I}_{6} & 0\end{array}
\right)
\eeq
as the invariant metric to raise and lower SO(6,6) indices. If we further split the index $M$ as $\,_M=( _a \,, \, _i \, , \, ^a \,, \, ^i)\,$, then the fluxes/ET dictionary is presented in Table~\ref{Table:ET_fluxes}. Notice the presence of electric $(\alpha=+)$ as well as magnetic $(\alpha=-)$ components within the embedding tensor $\,f_{\alpha MNP}={f_{\alpha MN}}^{Q} \, \eta_{QP}\,$. Both are simultaneously required in order to avoid a runaway behaviour for the dilaton modulus \cite{deRoo:1985jh}.

\begin{table}[t!]
\renewcommand{\arraystretch}{1.35}
\begin{center}
\begin{tabular}{|c|c|c|c|}
\hline
M-theory origin & Type IIA origin & Fluxes & Embedding tensor \\
\hline
\hline
${\omega_{b c}}^{a}$ & ${\omega_{b c}}^{a}$ & 
$\tilde{c}_{1}^{\,(I)}$ &  $f_{+\phantom{bc} a}^{\phantom{+}bc}$\\
\hline
${\omega_{a j}}^{k}$ & ${\omega_{a j}}^{k}$ & 
$\hat{c}_{1}^{\,(I)}$ & $f_{+\phantom{aj} k}^{\phantom{+}aj}$ \\
\hline
${\omega_{ka}}^{j}$ & ${\omega_{ka}}^{j}$ & 
$\check{c}_{1}^{\,(I)}$ & $f_{+\phantom{ka} j}^{\phantom{+}ka}$ \\
\hline
${\omega_{jk}}^{a}$ & ${\omega_{jk}}^{a}$ &  
$ b_{1}^{\,(I)}$  &  ${f_{-}}^{ibc}$ \\
\hline
$-{\omega_{ai}}^{7}$ & $ F_{ai}$ & 
$a_{2}^{\,(I)}$ & $-{f_{+}}^{ajk}$  \\
\hline
$-{\omega_{7i}}^{a}$ & non-geometric & 
$d_{0}^{\,(I)}$ &  $f_{-\phantom{bc} \,i}^{\phantom{-}bc}$ \\
\hline
$-{\omega_{a7}}^{i}$ & non-geometric & 
$c_{3}'^{\,(I)}$ &  ${f_{+jk}}^{a}$\\
\hline
\hline
$-\frac{1}{2} \, G_{aibj}$ & $-F_{aibj}$ & 
$a_{1}^{\,(I)}$ &  ${f_{+}}^{abk}$ \\
\hline
$\frac{1}{2} \,G_{ijk7}$ & $H_{ijk}$ & 
$b_{0}$ &  $- {f_{-}}^{abc}$\\
\hline
$\frac{1}{2} \,G_{ibc7}$ & $H_{ibc}$ & 
$c_{0}^{\,(I)}$ &  $f_{+\phantom{bc} \, i}^{\phantom{+}bc}$ \\
\hline
\hline
$\frac{1}{4} \,G_{aibjck7}$ & $F_{aibjck}$ & 
$a_{0}$ &  $- {f_{+}}^{abc}$ \\
\hline
\hline
non-geometric & $-F_{(0)}\,\,\,$ (Romans mass) & 
$a_{3}$ & ${f_{+}}^{ijk}$ \\
\hline
\end{tabular}
\end{center}
\caption{M-theory/type IIA fluxes and embedding tensor.}
\label{Table:ET_fluxes}
\end{table}

The consistency of a gauging in $\,\mathcal{N}=4\,$ supergravity \cite{Schon:2006kz} imposes a set of quadratic constraints on the embedding tensor $\,f_{\alpha MNP}\,$. These are given by
\beq
\label{QC_N=4}
f_{\alpha R[MN} \,\, {f_{\beta PQ]}}^{R} = 0
\hspace{10mm} \textrm{ and } \hspace{10mm}
\epsilon^{\alpha \beta} \, f_{\alpha MNR} \,\, {f_{\beta PQ}}^{R} = 0
\eeq 
where $\,\epsilon^{\alpha \beta}=\epsilon_{\alpha \beta}\,$ with $\,\epsilon^{+-}=-\epsilon^{-+}=1\,$ is used to raise and lower the SL(2) index $\alpha$. In order to make contact with the Scherk-Schwarz conditions in (\ref{SS_Constraints}) for M-theory reductions, we have to set the Romans mass to zero, \textit{i.e.} $a_{3}=0$, among the fluxes in Table~\ref{Table:ET_fluxes} as it corresponds to a non-geometric flux in M-theory. The explicit computation of the constraints in (\ref{QC_N=4}) produces the following conditions
\beq
\label{QCN4piece12}
\begin{array}{rcclc}
f_{\alpha R[MN} \,\, {f_{\beta PQ]}}^{R} = 0 & \rightarrow & & \textrm{Conditions $\,\,i)\,\,$ , $\,\,iii)\,\,$ and $\,\,v)\,\,$ in (\ref{ww_constraint})} & , \\[2mm]
\epsilon^{\alpha \beta} \, f_{\alpha MNR} \,\, {f_{\beta PQ}}^{R} = 0 & \rightarrow & & \textrm{Conditions $\,\,ii)\,\,$ and $\,\,vi)\,\,$ in (\ref{ww_constraint})} & .
\end{array}
\eeq
As a result, the quadratic constraints of $\mathcal{N}=4$ supergravity (\ref{QC_N=4}) fail to reproduce the two additional conditions $\,iv)\,$ and $\,vii)\,$ in (\ref{ww_constraint}). Therefore, demanding $\mathcal{N}=4$ in the effective theory is less restrictive than demanding a higher-dimensional interpretation as an ordinary Scherk-Schwarz reduction of M-theory. 

In ref.~\cite{Dibitetto:2011eu} it was shown that the $\,\mathcal{N}=4\,$ constraints (\ref{QC_N=4}) must be supplemented with two additional ones
\beq
\label{QC_N=8extra}
\left. \epsilon^{\alpha \beta} \, f_{\alpha [MNP} \,\, f_{\beta QRS]} \right|_{\textrm{SD}} = 0
\hspace{10mm} \textrm{ and } \hspace{10mm}
 f_{\alpha MNP} \,\, {f_{\beta}}^{MNP} = 0
\eeq 
in order to have an $\,\mathcal{N}=4 \rightarrow \mathcal{N}=8\,$ supersymmetry enhancement in the effective action. The label SD in the first constraint in (\ref{QC_N=8extra}) restricts it to the self-dual part of the SO(6,6) six-form $\,\epsilon^{\alpha \beta} \, f_{\alpha [MNP} \,\, f_{\beta QRS]}$. Once more, an explicit computation of these two constraints produces
\beq
\label{QCN8extrapiece12}
\begin{array}{rcclc}
\left. \epsilon^{\alpha \beta} \, f_{\alpha [MNP} \,\, f_{\beta QRS]} \right|_{\textrm{SD}} = 0 & \rightarrow & & \textrm{Conditions $\,\,iv)\,\,$ and $\,\,vii)\,\,$ in (\ref{ww_constraint})} & , \\[2mm]
f_{\alpha MNP} \,\, {f_{\beta}}^{MNP} = 0 & \rightarrow & & \textrm{No additional conditions} & ,
\end{array}
\eeq
hence completing the set of conditions in (\ref{ww_constraint}). In other words, there is a one-to-one correspondence between the $\mathcal{N}=8$ quadratic constraints and the conditions required by an ordinary Scherk-Schwarz reduction of M-theory.

\subsection{KK6 monopoles and $\mathcal{N}=8 \rightarrow  \mathcal{N}=4$ breaking}

In the previous section we have seen that requiring an $\mathcal{N}=4$ description of the effective supergravity allows for a relaxation of the conditions $\,iv)\,$ and $\,vii)\,$ in (\ref{ww_constraint}). However these still have to be imposed in any ordinary Scherk-Schwarz reduction of M-theory establishing the link to $\,\mathcal{N}=8\,$ supergravity.

On the other hand, a violation of some of the $\omega \, \omega=0$ conditions in (\ref{SS_Constraints}) has been connected to the presence of KK6 monopoles in the compactification scheme, thus going beyond twisted tori \cite{Villadoro:2007yq}. From the effective field theory point of view, we will refer to the would-be companion sources carrying negative charge as KKO6-planes following a similar terminology to that of ref.~\cite{Villadoro:2007yq}. Schematically,
\beq
{\omega_{[\bullet \bullet}}^{D} \, {\omega_{\bullet] D}}^{\psi}  \neq 0  \,\, \Rightarrow \,\, \textrm{Non-vanishing KK6 (KKO6) charge} \ ,
\eeq
where $\,\psi\,$ refers to the $S^{1}$ direction along which the KK6 is fibered and $\,[\bullet \bullet \bullet]\,$ specifies the 3-form dual to the 7-cycle filled by the KK6 and the $S^{1}$ fiber. The KK6 monopoles will induce a positive contribution to the scalar potential whereas the one coming from the KKO6-planes will be negative \cite{Villadoro:2007yq}.

In the case of $X_{7}=\mathbb{T}^7/(\mathbb{Z}_{2} \times \mathbb{Z}_{2} \times \mathbb{Z}_{2})$, there are 28 different KK6 monopoles compatible with the orbifold symmetries. These KK6's can be grouped as $6+6+3+1+3+6+3=28$ and source the r.h.s of the set of conditions in (\ref{ww_constraint}). KK6 monopoles in M-theory sourcing the $6+6$ conditions $\,i)\,$ and $\,ii)\,$ give rise to KK5's (fibered over $\eta^{i}$) and $\widetilde{\textrm{KK5}}$'s (fibered over $\eta^{a}$) monopoles in type IIA upon dimensional reduction. Those fibering $\eta^{7}$ source the $3+1$ conditions $\,iii)\,$ and $\,iv)\,$ and give rise to D6$_{\perp}$'s (threading $3$-cycles $\eta^{ajk}$) and D6$_{\parallel}$'s  (threading the $3$-cycle $\eta^{abc}$) upon reduction to type IIA along the $\eta^{7}$ direction. There are also $\,3+6+3\,$ KK6 monopoles sourcing the conditions $\,v)\,$, $\,vi)\,$ and $\,vii)\,$ which do not have an interpretation as type IIA sources. We denote them KK6$_{\perp}$ 's and $\,\widetilde{\textrm{KK6}}_{\perp}$'s (threading $3$-cycles $\eta^{ajk}$ and respectively fibered over $\eta^{i}$ and $\eta^{a}$) as well as KK6$_{\parallel}$'s (threading the $3$-cycle $\eta^{abc}$ and fibered over $\eta^{i}$). By looking at the conditions in (\ref{ww_constraint}), a non-vanishing net charge of KK6$_{\perp}$'s, $\,\widetilde{\textrm{KK6}}_{\perp}$'s and KK6$_{\parallel}$'s requires a non-trivial background for the fluxes $(c_{3}'^{(I)},d_{0}^{I})$. These are the M-theory fluxes without a type IIA counterpart in Table~\ref{Table:IIA_fluxes}, thus corresponding to non-geometric type IIA flux backgrounds. For the set of conditions in (\ref{ww_constraint}), the corresponding types of KK6 monopoles are summarised in Table~\ref{Table:KK6}.

\begin{table}[t!]
\renewcommand{\arraystretch}{1.25}
\begin{center}
\begin{tabular}{|c||c|c|c|c||c|c||c|c||c|c||c||c|c|}
\hline
Type & $x^{0}$ & $x^{1}$ & $x^{2}$ & $x^{3}$ & $\eta^{a}$ & $\eta^{i}$& $\eta^{b}$& $\eta^{j}$& $\eta^{c}$& $\eta^{k}$& $\eta^{7}$ & KK6 $\rightarrow$ type IIA & $\mathcal{N}=4$ ? \\
\hline
\hline
$i)$ & $\times$ & $\times$ & $\times$ & $\times$ & $\times$ & $\times$ &  & $\psi$ &  && $\times$ &  KK5 (KKO5) & no \\
\hline
$ii)$ & $\times$ & $\times$ & $\times$ & $\times$ & $\times$ & $\times$ & $\psi$ &  &  &  & $\times$ & $\widetilde{\textrm{KK5}}$ ($\widetilde{\textrm{KKO5}}$)  &  no \\
\hline
\hline
$iii)$ & $\times$ & $\times$ & $\times$ & $\times$ & $\times$ &  &  & $\times$ &  & $\times$ & $\psi$ & D6$_{\perp}$ (O6$_{\perp}$) & no  \\
\hline
$iv)$ & $\times$ & $\times$ & $\times$ & $\times$ & $\times$ &  & $\times$ &   & $\times$ &  & $\psi$ & D6$_{\parallel}$ (O6$_{\parallel}$) &  yes \\
\hline
\hline
$v)$ & $\times$ & $\times$ & $\times$ & $\times$ & $\times$ & $\psi$ &  & $\times$ &  & $\times$  &   &   KK6$_{\perp}$ (KKO6$_{\perp}$)  & no  \\
\hline
$vi)$ & $\times$ & $\times$ & $\times$ & $\times$ & $\times$ &  &  $\psi$ & $\times$ &  & $\times$  &  & $\widetilde{\textrm{KK6}}_{\perp}$ ($\widetilde{\textrm{KKO6}}_{\perp}$) &  no \\
\hline
$vii)$ & $\times$ & $\times$ & $\times$ & $\times$ & $\times$ &  & $\times$ &   & $\times$ & $\psi$ &  & KK6$_{\parallel}$ (KKO6$_{\parallel}$)  &  yes  \\
\hline
\end{tabular}
\end{center}
\caption{Set of KK6 (KKO6) monopoles compatible with the $X_{7}=\mathbb{T}^7/(\mathbb{Z}_{2} \times \mathbb{Z}_{2} \times \mathbb{Z}_{2})$ orbifold. They respectively source the r.h.s of the set of conditions in (\ref{ww_constraint}). Only D6$_{\parallel}$ (O6$_{\parallel}$) and KK6$_{\parallel}$ (KKO6$_{\parallel}$) sources can be consistently introduced in a background preserving $\,\mathcal{N}=4\,$ supersymmetry in four dimensions.}
\label{Table:KK6}
\end{table}

Our last concern is that of supersymmetry breaking in the presence of KK6 monopoles. From the general discussion of quadratic constraints in $\mathcal{N}=4,8$ supergravity of the previous section,  the effective theory preserves $\mathcal{N}=8$ supersymmetry only if no KK6 net charge is induced by the M-theory flux backgrounds. In this case the full set of conditions in ($\ref{QC_N=4}$) and ($\ref{QC_N=8extra}$) are satisfied implying an ordinary Scherk-Schwarz reduction of M-theory with no violation of the  constraints ($\ref{SS_Constraints}$). If the M-theory background fluxes induce a non-vanishing charge for KK6 (KKO6) monopoles corresponding to D6$_{\parallel}$ (O6$_{\parallel}$), KK6$_{\parallel}$ (KKO6$_{\parallel}$) or both, then $\,\mathcal{N}=4\,$ supersymmetry is still preserved but one goes beyond Scherk-Schwarz reductions of M-theory due to the violation of (\ref{SS_Constraints}). We will exhaustively explore these two types of effective theories in the next section.

\subsection{Universal IIA moduli, KK6 monopoles and scalar potential}
\label{sec:scalings}

A way of understanding the effect of including M-theory sources in the background is to analyse the moduli powers appearing in the scalar potential. In order to make contact with previous results in the literature \cite{Hertzberg:2007wc,Silverstein:2007ac,Haque:2008jz,Caviezel:2008tf,Danielsson:2009ff} we will reinterpret the M-theory potential from a type IIA point of view. To this end, let us introduce the three universal IIA moduli fields $(\tau,\rho,\sigma)$ entering the 10d metric
\beq
ds_{10}^2 = \tau^{-2} \, ds_{4}^2 + \rho \, ( \, \sigma^{-3} \,  M_{ab} \, dy^{a} dy^{b} + \sigma^{3} \,  M_{ij} \, dy^{i} dy^{j} \, ) \ ,
\eeq
which are related to the STU fields as
\beq
\tau = \textrm{Im}(S)^{1/4} \, \textrm{Im}(T)^{3/4}
\hspace{3mm} , \hspace{3mm}
\rho = \textrm{Im}(U)
\hspace{3mm} , \hspace{3mm}
\sigma = \textrm{Im}(S)^{-1/6} \, \textrm{Im}(T)^{1/6} \ .
\eeq 
We follow the conventions in appendix B of ref.~\cite{Dibitetto:2014sfa} regarding dimensional reduction of 10d type IIA supergravity.

Setting the axions to zero, namely $\textrm{Re}(S)=\textrm{Re}(T)=\textrm{Re}(U)=0$, the computation of the scalar potential from the  M-theory superpotential (\ref{Superpotential_Flux_MTheory_ISO}) reveals the following $\tau$-dependence structure
\beq
\label{V_M-theory}
V_{\textrm{M-theory}}(\tau,\rho,\sigma) = \frac{1}{32}  \, \displaystyle\sum_{n=0}^{4} V_{n}(\tau,\rho,\sigma)  = \frac{1}{32} \, \displaystyle\sum_{n=0}^{4} A_{n}(\rho,\sigma) \, \tau^{-n} \ .
\eeq
The functions $\,A_{n}(\rho,\sigma)\,$ that determine the different terms $\,V_{n}=A_{n}(\rho,\sigma) \, \tau^{-n}\,$ in the potential take the following form:
\beq
\begin{array}{llll}
A_{0}(\rho,\sigma) & = & 3 \, \rho^{-3} \, (c_3' \, \sigma^3 - d_0 \, \sigma^{-3} )^2  \\[3mm]
A_{1}(\rho,\sigma) & = & 6 \, \rho^{-2} \, [\, (2 c_1-\tilde{c}_{1}) \,  c_3' \, \sigma^{9/2} + 2 \, ( b_1 \, c_3' + (c_1 -  \tilde{c}_{1}) \, d_0 )\, \sigma^{-3/2}  + ( b_1 \, c_3' +2 \,  c_1 \, d_0 ) \, \sigma^{-3/2}  \, ] \\[3mm]
A_{2}(\rho,\sigma) & = & \rho^{-3} \, (b_0^2 \, \sigma^{-9}+3 \, c_0^2 \, \sigma^3) \\[2mm]
&+& 3\,  \rho^{-1} [\,  b_1^2 \, \sigma^{-9} - 4 \, b_1  ( 2 c_1-\tilde{c}_{1}) \, \sigma^{-3}  - (2 c_1 - \tilde{c}_{1}  )^2  \,\sigma^3 \, ] &  \\[2mm]
&+& 18 \,  \rho^{-1} \, a_{2} \, ( c_3'  \,\sigma^3 + d_0  \, \sigma^{-3} )  \\[3mm]
A_{3}(\rho,\sigma) & = & - 6 \, (2 c_1-\tilde{c}_{1} ) \, a_2 \, \sigma ^{3/2} - 6 \, a_2 \, b_1 \, \sigma^{-9/2}  \\[3mm]
A_{4}(\rho,\sigma) & = & a_0^2 \, \rho^{-3} + 3 \, a_1^2 \, \rho^{-1}+3 \, a_2^2 \, \rho \ .
\end{array}
\eeq
Let us discuss the $V_{n}$ terms in the M-theory scalar potential (\ref{V_M-theory}) when adopting a type IIA point of view using the M-theory/type IIA dictionary in Table~\ref{Table:ET_fluxes}. Recall that only the flux parameters $(c_{3}',d_{0})$ are genuine M-theory metric fluxes without type IIA counterparts. These fluxes are responsible for the two terms $V_{0}\propto \tau^{0} \rho^{-3}$ and $V_{1}\propto \tau^{-1} \rho^{-2}$ which have no analogous in a regular IIA orientifold model \cite{Hertzberg:2007wc,Caviezel:2008tf,Danielsson:2009ff} thus corresponding to \textit{non-geometric} contributions in a IIA incarnation of the potential (\ref{V_M-theory}). The three pieces inside $V_{1}$ account for the net charge of KK6$_{\perp}$ (KKO6$_{\perp}$), $\widetilde{\textrm{KK6}}_{\perp}$ ($\widetilde{\textrm{KKO6}}_{\perp}$) and KK6$_\parallel$ (KKO6$_\parallel$) monopoles, respectively. The  term $V_{2}$ plays a central role in stabilising moduli and contains two types of contributions proportional to $\tau^{-2} \rho^{-3}$ and $\tau^{-2} \rho^{-1}$ respectively: the former is sourced by NS-NS fluxes $H_{(3)}$ in the IIA picture (first line in $A_{2}$) whereas the latter is induced by metric IIA fluxes $\omega_{(\textrm{IIA})}$ (second line in $A_{2}$) as well as the two fluxes $(c_{3}',d_{0})$ which are \textit{non-geometric} in the IIA description of the STU-model (third line in $A_{2}$). The two pieces in $V_{3} \propto \tau^{-3} \rho^{0}$ respectively account for the net charge of D6$_{\perp}$ and D6$_{\parallel}$ sources and the corresponding orientifold planes. Finally the $V_{4}$ term contains the type IIA R-R contributions to the scalar potential. Notice the absence of the Romans contribution $V_{\textrm{Romans}} \propto \tau^{-4} \rho^{3}$ that would be induced by the flux parameter $a_{3}$ which is not present in the M-theory setup and played a central role in the construction of ref.~\cite{Haque:2008jz} producing de Sitter (dS$_{4}$) solutions.

In the previous section we saw that including a net charge for those KK6 sources in M-theory which correspond to KK5 and $\widetilde{\textrm{KK5}}$ monopoles in the IIA picture -- types $i)$ and $ii)$ in Table~\ref{Table:KK6} -- was not compatible with preserving $\,\mathcal{N}=4\,$ supersymmetry in the effective action. The reason was that the associated conditions $i)$ and $ii)$ in (\ref{ww_constraint_iso}) still hold after relaxing $\mathcal{N}=8 \rightarrow \mathcal{N}=4$. The effect of adding such monopoles has been investigated in refs~\cite{Silverstein:2007ac,Haque:2008jz} and found to induce an extra piece $V_{\textrm{KK5}} \propto \tau^{-2} \rho^{-1}$ in the potential supplementing the one already induced by the IIA metric flux $V_{\omega_{\textrm{IIA}}} \propto \tau^{-2} \rho^{-1}$ with the same moduli powers. More importantly, this extra piece $V_{\textrm{KK5}}$ turned out to help in finding de Sitter solutions \cite{Silverstein:2007ac,Haque:2008jz}. Even though we cannot include such KK5 and $\widetilde{\textrm{KK5}}$ monopoles when demanding $\,\mathcal{N}=4\,$ supersymmetry, the M-theory fluxes $(c_{3}',d_{0})$ will potentially induce the desired $\tau^{-2} \rho^{-1}$ extra piece within $V_{2}$ (third line in $A_{2}$)\footnote{This $\,\tau^{-2} \rho^{-1}\,$ extra piece within $V_{2}$ can be obtained from the M-theory superpotential (\ref{Superpotential_Flux_MTheory_ISO}) but not from the type IIA superpotential (\ref{Superpotential_Flux_IIA_ISO}) due to the lack of the two relevant fluxes $(c_{3}',d_{0})$. This result also holds after turning on the three STU axions as they do not modify the second and third lines in $A_{2}(\rho,\sigma)$. Only $A_{4}(\rho,\sigma)$ and the first line in $A_{2}(\rho,\sigma)$ corresponding to R-R and NS-NS gauge fluxes in the IIA picture are modified by the STU axions.}. Despite this promising fact, only Anti-de Sitter (AdS$_{4}$) solutions will happen to exist in these $\,\mathcal{N}=4\,$ STU-models.

\section{Taxonomy of M-theory flux vacua}
\label{sec:taxonomy}

In this section we will exhaustively classify the entire set of critical points of the scalar potential obtained from the M-theory superpotential (\ref{Superpotential_Flux_MTheory_ISO}) when demanding $\,\mathcal{N}=4\,$ supersymmetry in the effective action. In addition to the superpotential analysis, we have also verified all the results by explicit computations using the $\,\mathcal{N}=4\,$ scalar potential directly built from the embedding tensor \cite{Schon:2006kz}.

\subsection{Exploiting dualities in the effective theory}

The $\,\mathcal{N}=1\,$ supergravity model we derived in section~\ref{section:M-Theory_effective} can be formally viewed as a discrete $\,\mathbb{Z}_{2} \times \mathbb{Z}_{2}\,$ truncation of the $\,\mathcal{M}^{\mathcal{N}=4}_{\textrm{scalar}}=[\textrm{SL}(2)/\textrm{SO}(2)]  \times [\textrm{SO}(6,6)/\textrm{SO}(6) \times \textrm{SO}(6)]\,$ coset space spanned by the $2+36$ scalar fields of the $\mathcal{N}=4$ theory. The seven complex moduli $T_{A}=(S,T_{I},U_{I})$ correspond to the seven dilatons (Cartan generators) as well as seven axions (positive roots), and span the K\"ahler manifold $\mathcal{M}^{\textrm{(non-iso)}}_{\textrm{scalar}}=[\textrm{SU}(1,1)/\textrm{U}(1)]^{7}$. As we already discussed, demanding \textit{isotropy} imposes an additional ``plane exchange" SO(3) cyclic symmetry among the three two-tori in $\mathbb{T}^{6}=\mathbb{T}^{2} \times \mathbb{T}^{2} \times \mathbb{T}^{2}$ (inside $X_{7}$). This additional symmetry can be interpreted as an enhancement of the truncation from a $\mathbb{Z}_{2} \times \mathbb{Z}_{2}$ truncation to an $\textrm{SO}(3)$ truncation. After taking into account the isotropic identifications in (\ref{LimIsoFields}), the three complex moduli in the STU-model serve as coordinates in the K\"ahler submanifold $\mathcal{M}^{\textrm{(iso)}}_{\textrm{scalar}}=[\textrm{SU}(1,1)/\textrm{U}(1)]^{3} \subset [\textrm{SU}(1,1)/\textrm{U}(1)]^{7}$.

The $\,\mathcal{M}^{\textrm{(iso)}}_{\textrm{scalar}}\,$ coset space is a symmetric space and therefore any point can be connected to any other via a non-compact $\textrm{SU}(1,1)^{3}$ transformation. The action of this transformation on the STU-model is that of rescaling and shifting the moduli fields as
\beq
\label{non-compact_transf_ISO}
S \rightarrow \lambda_{S} \, S + \Delta_{S}
\hspace{6mm} ,  \hspace{6mm} 
T \rightarrow \lambda_{T} \, T + \Delta_{T}
\hspace{6mm} ,  \hspace{6mm} 
U \rightarrow \lambda_{U} \, U + \Delta_{U} \ ,
\eeq
with $\lambda_{S,T,U}$ and $\Delta_{S,T,U}$ being real parameters. By using the tranformations (\ref{non-compact_transf_ISO}), any moduli configuration corresponding to a critical point of the scalar potential can be brought to the \textit{origin} of the moduli space defined as
\beq
\label{origin_moduli}
S_{0} = T_{0} = U_{0} = i \ .
\eeq
After bringing the moduli configuration to the origin (\ref{origin_moduli}), the associated flux parameters entering the superpotential (\ref{Superpotential_Flux_MTheory_ISO}) will change in order to leave the scalar potential invariant. Notice that the M-theory superpotential in (\ref{Superpotential_Flux_MTheory_ISO}) is not only quadratic on the moduli but it also contains the linear dependences as well as the constant term. This ensures that the new flux background obtained after bringing the moduli configuration to the origin will not lie outside the family of STU-models we are considering here. In other words, the M-theory backgrounds form a closed set under the action of the duality tranformations (\ref{non-compact_transf_ISO}).

The above argument allows us to look for moduli stabilisation at any point in moduli space and, for the sake of simplicity, we will choose such a point to be the origin (\ref{origin_moduli}). This does not imply any loss of generality as long as one keeps the complete set of flux parameters in the superpotential (\ref{Superpotential_Flux_MTheory_ISO}). Focusing on the M-theory backgrounds preserving at least $\mathcal{N}=4$ supersymmetry in the effective action, the structure of critical points can therefore be obtained by solving the algebraic system
\beq
\label{ideal_solutions}
\left\langle  \textrm{(relaxed) set of conditions in } (\ref{ww_constraint_iso}) \,\,\, , \,\,\, \left. \frac{\partial V}{\partial \phi}\right|_{S_{0}=T_{0}=U_{0}=i} \right\rangle = 0 \ ,
\eeq
which consists of quadratic conditions on the flux parameters. As discussed in the previous section, the conditions $\,iv)\,$ and $\,vii)\,$ in (\ref{ww_constraint_iso}) can be consistently  relaxed if one goes beyond twisted tori reductions but still requires the effective action to have $\mathcal{N}=4$ supersymmetry. The above set of equations in (\ref{ideal_solutions}) can be completely solved -- with or without relaxing $\,iv)\,$ and $\,vii)\,$ -- by using algebraic geometry tools included in the computational package \mbox{\textsc{singular}} \cite{DGPS}. In particular, we have used the GTZ built-in algorithm for primary decomposition of ideals. The outcome is that (\ref{ideal_solutions}) contains several prime factors, each of which corresponds to a physically different content of KK6 monopoles in the M-theory background.  We will discuss them in detail later on.

Some other advantages of bringing the moduli configurations to the origin are : $i)$ closed expressions for the particle mass spectra at a critical point of the scalar potential have been worked out \cite{Borghese:2010ei} $\,\,ii)$ the fermion mass terms get a much simpler form. The fermion masses can be viewed as ``dressing up" the embedding tensor with the moduli dependence \cite{Schon:2006kz}. When evaluated at the origin, the $\mathcal{N}=4$ gravitini mass matrix acquires the very simple form
\beq
\label{gravitini_mass_matrix}
\mathcal{A} = -\frac{3}{8 \sqrt{2}} \, 
\left(
\begin{array}{cc}
\mathcal{A}_{1} & 0 \\
0 & \mathcal{A}_{2} \, \mathbb{I}_{3\times 3}
\end{array}
\right) \ ,
\eeq
with the two independent entries given by
\beq
\begin{array}{cclc}
\mathcal{A}_{1}  &=&  (\, a_{0} - 3 a_{2} - 3 b_{1} - 3 \, (2 c_{1} - \tilde{c}_{1})  + 3 c_{3}' + 3 d_{0} \, ) + i \, ( \,3 c_{0} -3 a_{1} - b_{0}  \, ) & ,\\[2mm]
\mathcal{A}_{2}  &=&  (\, a_{0} + a_{2} + b_{1} + 2 c_{1} + 3 \tilde{c}_{1}  - c_{3}' - d_{0} \, ) + i \, ( \,a_{1} - b_{0} - c_{0} \, ) & .
\end{array}
\eeq
Notice that $\,\mathcal{A}_{1}= W^{\textrm{(iso)}}_{\textrm{M-Theory}}\,$ at the origin of the moduli space as it has to in order to identify the $\,\mathcal{N}=1\,$ gravitino mass with $\,|W^{\textrm{(iso)}}_{\textrm{M-Theory}}|$. The gravitino mass matrix in (\ref{gravitini_mass_matrix}) can be used to determine the amount of supersymmetry preserved at a critical point of the scalar potential. Provided an AdS$_{4}$ vacuum solution, it will preserve $\mathcal{N}=1$ supersymmetry if ${|\mathcal{A}_{1}|^2=-3 V_{0}}$ and $|\mathcal{A}_{1}|^2 \neq |\mathcal{A}_{2}|^2$. Similarly, it will preserve $\mathcal{N}=3$ supersymmetry if ${|\mathcal{A}_{2}|^2=-3 V_{0}}$ and $|\mathcal{A}_{1}|^2 \neq |\mathcal{A}_{2}|^2$. Finally it will preserve $\mathcal{N}=4$ supersymmetry if ${|\mathcal{A}_{1}|^2=|\mathcal{A}_{2}|^2=-3 V_{0}}\,$ and will be non-supersymmetric otherwise.

\subsection{Backgrounds without KK6 (KKO6)}

Let us start by studying the case of not having KK6 monopoles of any type. Therefore, the full set of conditions in (\ref{ww_constraint_iso}) have to be imposed and maximal $\,\mathcal{N}=8\,$ supersymmetry is preserved in the effective action. The M-theory flux background solving (\ref{ideal_solutions}) in this case is given by
\beq
\label{vacuum0_fluxes}
a_{0} = a_{1} = a_{2} = b_{0} =  b_{1} = c_{0} =  c_{1} = \tilde{c}_{1} = 0  
\hspace{10mm} \textrm{ and } \hspace{10mm}
c_{3}' = d_{0} = \lambda \ ,
\eeq
so that the M-theory superpotential in (\ref{Superpotential_Flux_MTheory_ISO}) takes the simple form
\beq
\label{vacuum0_superpotential}
W^{\textrm{(iso)}}_{\textrm{M-Theory}} = -3 \, \lambda \, T \, (S+T) \ .
\eeq
Within the class of $\,\mathcal{N}=1\,$ STU-models, this superpotential specifies a no-scale supergravity so the vacuum corresponds to a non-supersymmetric and Minkowski ($V_{0}=0$) critical point \textit{with} flat directions. The analysis of the torsion classes shows non-vanishing $\,\widetilde{W}_{1}\,$ and $\,\widetilde{W}_{27}$, thus specifying a general co-calibrated $G_{2}$-structure (\ref{G2_structure}). 

The associated M-theory background only contains non-geometric fluxes $\,(c_{3}' , d_{0})\,$ in a type IIA incarnation. This is compatible with the results found in refs~\cite{Derendinger:2004jn,Camara:2005dc,Dall'Agata:2009gv,Dibitetto:2012ia} regarding type IIA moduli stabilisation based on the superpotential (\ref{Superpotential_Flux_IIA_ISO}). In ref.~\cite{Dibitetto:2012ia} an exhaustive classification of vacua compatible with the full set (\ref{ww_constraint_iso}) of $\,\mathcal{N}=8\,$ constraints (after setting $c_{3}' = d_{0} = 0$) showed the necessity of a non-vanishing Romans mass ($a_{3} \neq 0$) in order to achieve \textit{full} moduli stabilisation. The Romans flux parameter in type IIA does not descend directly from \mbox{M-theory} (see Table~\ref{Table:ET_fluxes}), so the IIA solutions in ref.~\cite{Dibitetto:2012ia} will not appear in an M-theory context. Finally the M-theory $\omega$-twist corresponds to $G_{\omega}=\textrm{Solv}_{6} \rtimes \textrm{U}(1)$ in agreement with the analysis of twist groups performed in ref.~\cite{DallAgata:2005fm}.

\subsection{Backgrounds with KK6 (KKO6)}
\label{sec:N=4_gauging}

We have rederived the result that there is no moduli stabilisation (without flat directions) in the absence of KK6 (KKO6) monopoles \cite{DallAgata:2005fm}. Next step is then to remove the conditions
\beq
\label{QC4_relax}
\begin{array}{rrlrllllll}
iv)  \hspace{10mm} & {\omega_{[ij}}^{D} \, {\omega_{k]D}}^{7}  \,\,=\,\, 0 &  \hspace{3mm} \rightarrow \hspace{3mm} &   3\,  b_{1}\, a_{2}    =  0 & ,  \\[2mm]
vii) \hspace{10mm} & {\omega_{[7i}}^{D} \, {\omega_{j]D}}^{k}  \,\,=\,\, 0 &  \hspace{3mm} \rightarrow \hspace{3mm} &   b_{1} c_{3}' + 2 \, c_{1} \,  d_{0} =  0 &   ,
\end{array}
\eeq
from the system (\ref{ww_constraint_iso}) in order to preserve only $\mathcal{N}=4$ and investigate the physical implications. Running the primary decomposition algorithm for the relaxed algebraic system in (\ref{ideal_solutions}) one finds three prime factors they all of dimension one. We will discuss each of them separately and show how \textit{full} moduli stabilisation can take place in M-theory backgrounds containing KK6 (KKO6) monopoles.

\subsubsection{Including only KK6 (KKO6) $\rightarrow$ D6$_{\parallel}$ (O6$_{\parallel}$) sources}
\label{sec:only D6}

The first prime factor in the decomposition of (\ref{ideal_solutions}) is compatible with a relaxation of the condition
\beq
\label{D6_relax}
\begin{array}{rrlrllllll}
iv)  \hspace{10mm} & {\omega_{[ij}}^{D} \, {\omega_{k]D}}^{7}  \,\,\neq\,\, 0 &  \hspace{3mm} \rightarrow \hspace{3mm} &   3\,  b_{1}\, a_{2}    \neq  0 &  ,
\end{array}
\eeq
whereas the rest of conditions in (\ref{ww_constraint_iso}) are still satisfied. This case is then interpreted as an M-theory background which only includes those KK6 (KKO6) monopoles that can be interpreted as D6$_{\parallel}$ (O6$_{\parallel}$) type IIA sources upon reduction.  

By explicitly solving this prime factor we find a one dimensional family of M-theory flux backgrounds of the form
\beq
a_{0}  = b_{0} = a_{1}  = c_{0} =  c_{1} = \tilde{c}_{1} = c_{3}' = d_{0} = 0 
\hspace{10mm} \textrm{ and } \hspace{10mm}
b_{1} = a_{2} = \lambda \ .
\eeq
After substitution into (\ref{Superpotential_Flux_MTheory_ISO}),  the M-theory superpotential reads
\beq
W^{\textrm{(iso)}}_{\textrm{M-Theory}} = 3 \, \lambda \, U \, (S+U) \ ,
\eeq
corresponding to a no-scale STU-model analogous to that in (\ref{vacuum0_superpotential}) upon exchanging $T \leftrightarrow U$. The associated vacuum -- we will refer to it as ``\textbf{vac 0}"  from now on -- is a non-supersymmetric Minkowski vacuum with non-vanishing $\widetilde{W}_{1}$ and $\widetilde{W}_{27}$ torsion classes in (\ref{G2_structure}). 

Using the mass formula in ref.~\cite{Borghese:2010ei}, the scalar mass spectrum is given by
\beq
\label{vacuum0_scalars}
m^2 = \frac{9}{8} \, \lambda^2 \,\, (\times 1) \,\,\, , \,\,\, \frac{1}{2} \, \lambda^2 \,\, (\times 6) \,\,\, , \,\,\, \frac{1}{8} \, \lambda^2 \,\, (\times 9) \,\,\, , \,\,\, 0 \,\, (\times 22) \ ,
\eeq
so it does not contain tachyons but presents thirteen flat directions, \textit{i.e.}, zero-mass modes not associated to Goldstone bosons. The spectrum of vector masses reads 
\beq
\label{vacuum0_vectors}
m^2 = \frac{1}{2}\, \lambda^2 \,\, (\times 3) \,\,\, , \,\,\, \frac{1}{8}\, \lambda^2 \,\, (\times 6)  \,\,\, , \,\,\, 0 \,\, (\times 3) \ ,
\eeq
and contains three massless vectors reflecting the residual $G_{\textrm{res}}=\textrm{SO}(3)$ cyclic symmetry of the isotropic STU-model.

In a type IIA interpretation of this M-theory flux vacuum, we have introduced D6$_{\parallel}$/O6$_{\parallel}$ sources in the background wrapping the 3-cycle $\,\eta^{abc}\,$ in order to cancel a flux-induced tadpole for the R-R gauge potential $C_{(7)}$. The BI for $F_{(2)}$ along the internal space $X_{6}$ reads
\beq
dF_{(2)} = \omega F_{(2)} = N_{\textrm{O6$_{\parallel}$}} - N_{\textrm{D6$_{\parallel}$}} = 3 \, b_{1} \, a_{2} = 3 \, \lambda^2 > 0 \ ,
\eeq
thus demanding O6$_{\parallel}$ orientifold planes lifting to KKO6-planes in M-theory (see ref.~\cite{Villadoro:2007yq} and references therein for a discussion of the lifting).

\subsubsection{Including only KK6 (KKO6) $\rightarrow$ KK6$_{\parallel}$ (KKO6$_{\parallel}$)  sources}
\label{sec:KK6_parallel only}

The second prime factor in the decomposition of the algebraic system (\ref{ideal_solutions}) is compatible with relaxing
\beq
\label{KK6_relax}
\begin{array}{rrlrllllll}
vii) \hspace{10mm} & {\omega_{[7i}}^{D} \, {\omega_{j]D}}^{k}  \,\,\neq\,\, 0 &  \hspace{3mm} \rightarrow \hspace{3mm} &   b_{1} c_{3}' + 2 \, c_{1} \,  d_{0} \neq  0 &   ,
\end{array}
\eeq
but still requires the rest of the conditions in (\ref{ww_constraint_iso}) to vanish. Therefore, the resulting \mbox{M-theory} backgrounds only include KK6$_{\parallel}$ (KKO6$_{\parallel}$) monopoles. Backgrounds including a net charge of these objects do not admit a description in terms of ordinary type IIA orientifolds. Instead, they correspond to \textit{non-geometric} type IIA backgrounds.

Solving this prime factor explicitly reveals a rich structure of M-theory flux vacua they all compatible with
\beq
\label{fluxes_common_onlyKK6}
a_{2} = c_{3}' = 0 \,\,\,\, \textrm{ and } \,\,\,\, c_{1} = \tilde{c}_{1} = \lambda \ , 
\eeq
so that the $U^2$ and $T^2$ terms in the superpotential (\ref{Superpotential_Flux_MTheory_ISO}) are absent. Up to some discrete multiplicities there are eight inequivalent vacua we have denoted ``\textbf{vac 1}" to ``\textbf{vac 8}". The physical implications of these M-theory backgrounds are very diverse and we have carried out a detailed analysis in Appendix~\ref{app:only_KK6_parallel}. A brief summary of the main results is presented also in Table~\ref{Table:Landscape}. In all the solutions the net charge of KK6$_{\parallel}$ (KKO6$_{\parallel}$) sources is
\beq
N_{\textrm{KK6}_{\parallel}} - N_{\textrm{KKO6}_{\parallel}}=b_{1} \, c_{3}' + 2 \, c_{1} \,  d_{0}  < 0 \ ,
\eeq
hence requiring KKO6$_{\parallel}$-planes to be present in the background. At this point we want to highlight that \textit{full} moduli stabilisation at supersymmetric as well as at non-supersymmetric vacua is achieved for some of these M-theory backgrounds.

\begin{table}[t!]
\renewcommand{\arraystretch}{1.35}
\begin{center}
\scalebox{0.90}[0.90]{
\begin{tabular}{|c|c|c|c|c|c|c|c|c|c|}
\hline
ID & STU-model & D6$_{\parallel}$ (O6$_{\parallel}$) / KK6$_{\parallel}$ (KKO6$_{\parallel}$) & Stable & Flat dir. & SUSY & dim($G_{\textrm{res}})$ & $\widetilde{W}_{27}$ \\
\hline
\hline
\textbf{vac 0} & no-scale  & yes / no  & $\checkmark$  & yes & $\mathcal{N}=0$ & $3$  & $\neq 0$ \\  
\hline
\hline
\textbf{vac 1} & Model 1  & no  / yes  & $\checkmark$  & yes & $\mathcal{N}=3$ & $3$  & $\neq 0$ \\  
\hline
\textbf{vac 2} & Model 2  & no  / yes  & $\checkmark$  & yes & $\mathcal{N}=0$ & $3$  & $\neq 0$ \\  
\hline
\textbf{vac 3} & Model 1  & no  / yes  & $\checkmark$  & no & $\mathcal{N}=0$ & $3$  & $0$ \\  
\hline
\textbf{vac 4} & Model 2  & no  / yes  & $\checkmark$  & no & $\mathcal{N}=1$ & $3$  & $0$ \\  
\hline
\textbf{vac 5} & Model 1  & no  / yes  & $\checkmark$  & no & $\mathcal{N}=0$ & $3$  & $0$ \\  
\hline
\textbf{vac 6} & Model 2  & no  / yes  & $\times$  & no & $\mathcal{N}=0$ & $3$  & $\neq 0$ \\  
\hline
\textbf{vac 7} & Model 2  & no  / yes  & $\times$  & no & $\mathcal{N}=0$ & $3$  & $\neq 0$ \\  
\hline
\textbf{vac 8} & Model 2  & no  / yes  & $\checkmark$  & no & $\mathcal{N}=0$ & $3$  & $\neq 0$ \\  
\hline
\hline
\textbf{vac 9} & Model 3  & yes  / yes  & $\checkmark$  & yes & $\mathcal{N}=3$ & $6$  & $\neq 0$ \\  
\hline
\textbf{vac 10} & Model 4  & yes  / yes  & $\checkmark$  & no & $\mathcal{N}=0$ & $6$  & $\neq 0$ \\  
\hline
\textbf{vac 11} & Model 4  & yes  / yes  & $\checkmark$  & no & $\mathcal{N}=1$ & $6$  & $0$ \\  
\hline
\textbf{vac 12} & Model 3  & yes  / yes  & $\checkmark$  & no & $\mathcal{N}=0$ & $6$  & $0$ \\  
\hline
\textbf{vac 13} & Model 3  & yes  / yes  & $\times$  & no & $\mathcal{N}=0$ & $6$  & $0$ \\  
\hline
\textbf{vac 14} & Model 3  & yes  / yes  & $\times$  & no & $\mathcal{N}=0$ & $3$  & $\neq 0$ \\  
\hline
\textbf{vac 15} & Model 3  & yes  / yes  & $\times$  & no & $\mathcal{N}=0$ & $3$  & $\neq 0$ \\  
\hline
\textbf{vac 16} & Model 4  & yes  / yes  & $\times$  & no & $\mathcal{N}=0$ & $3$  & $\neq 0$ \\  
\hline
\textbf{vac 17} & Model 3  & yes  / yes  & $\times$  & no & $\mathcal{N}=0$ & $3$  & $\neq 0$ \\  
\hline
\end{tabular}
}
\end{center}
\caption{Data associated to the M-theory landscape compatible with KK6 (KKO6) sources preserving $\,\mathcal{N}=4\,$ supersymmetry in four dimensions. All the M-theory backgrounds happen to require a non-vanishing torsion class $\,\widetilde{W}_{1} \neq 0$.}
\label{Table:Landscape}
\end{table}

\subsubsection{Including both types of KK6 (KKO6) sources}
\label{sec:both_KK6}

The third prime factor in the decomposition of the system (\ref{ideal_solutions}) demands to simultaneously relax the two condtions
\beq
\label{D6&KK6_relax}
\begin{array}{rrlrllllll}
iv)  \hspace{10mm} & {\omega_{[ij}}^{D} \, {\omega_{k]D}}^{7}  \,\,\neq\,\, 0 &  \hspace{3mm} \rightarrow \hspace{3mm} &   3\,  b_{1}\, a_{2}    \neq  0 &  , \\[2mm]
vii) \hspace{10mm} & {\omega_{[7i}}^{D} \, {\omega_{j]D}}^{k}  \,\,\neq\,\, 0 &  \hspace{3mm} \rightarrow \hspace{3mm} &   b_{1} c_{3}' + 2 \, c_{1} \,  d_{0} \neq  0 &   ,
\end{array}
\eeq
but still imposes the rest of the conditions in (\ref{ww_constraint_iso}) in order to preserve $\,\mathcal{N}=4\,$ supersymmetry. The corresponding M-theory backgrounds then simultaneously include D6$_{\parallel}$ (O6$_{\parallel}$) as well as KK6$_{\parallel}$ (KKO6$_{\parallel}$) sources and cannot be interpreted as an ordinary type IIA orientifold. 

This last prime factor is also of dimension one and therefore can be solved explicitly. We find a rich structure of AdS$_{4}$ critical points  they all compatible with
\beq
\label{fluxes_common_bothKK6}
2 \, c_{1} = \tilde{c}_{1} = \lambda \ ,
\eeq
so that the $UT$ term in the superpotential (\ref{Superpotential_Flux_MTheory_ISO}) is absent. Up to discrete multiplicities, we now find nine inequivalent M-theory flux vacua labelled as ``\textbf{vac 9}"  to ``\textbf{vac 17}". As in the previous case the phenomenological consequences are very diverse and we have moved a detailed discussion of these M-theory backgrounds to Appendix~\ref{app:both_KK6}. A summary of the main features of these vacua is also included in Table~\ref{Table:Landscape}.  For these solutions the net charge of localised sources are
\beq
N_{\textrm{O6$_{\parallel}$}} - N_{\textrm{D6$_{\parallel}$}} = 3 \, b_{1} \, a_{2} > 0
\hspace{8mm} \textrm{ and } \hspace{8mm} 
N_{\textrm{KK6}_{\parallel}} - N_{\textrm{KKO6}_{\parallel}}=b_{1} \, c_{3}' + 2 \, c_{1} \,  d_{0}  < 0 \ ,
\eeq
requiring the presence of O6$_{\parallel}$- and $\textrm{KKO6}_{\parallel}$-planes in the backgrounds. Moduli fields can also be \textit{fully} stabilised at supersymmetric and non-supersymmetric vacua for some of these M-theory backgrounds but, generically, instabilities happen to occur more often.

\subsection{Monopoles and duality orbits of $\mathcal{N}=4$ gaugings}

The results in the previous section have shown that M-theory backgrounds including KK6 (KKO6) monopoles lead to moduli stabilisation in the effective STU-models. Now we will investigate the $\,\mathcal{N}=4\,$ gaugings underlying such M-theory backgrounds with sources. 

Let us start by recalling the $[\textrm{SU}(1,1)/\textrm{U}(1)]^3$ duality transformation in (\ref{non-compact_transf_ISO}) needed to bring a given moduli configuration to the origin, namely,
\beq
\label{non-compact_transf_ISO2}
S \rightarrow \lambda_{S} \, S + \Delta_{S}
\hspace{6mm} ,  \hspace{6mm} 
T \rightarrow \lambda_{T} \, T + \Delta_{T}
\hspace{6mm} ,  \hspace{6mm} 
U \rightarrow \lambda_{U} \, U + \Delta_{U} \ .
\eeq
The action of applying (\ref{non-compact_transf_ISO2}) to the M-theory superpotential in (\ref{Superpotential_Flux_MTheory_ISO}) has the effect of redefining the flux background (moduli couplings) in the following way
%
%
%
\beq
\label{Flux_redefinition}
\begin{array}{lll}
a_{0}  \rightarrow  a_{0} + 3 \Delta_U (b_1 \Delta_S + C_{1}\Delta_T - a_1)+3 a_2 \Delta_U^2- b_0 \Delta_S+3 \Delta_T (c_0-c_{3}' \Delta_T- d_0 \Delta_S) \\[2mm]
a_{1}  \rightarrow  \lambda_U (a_1-2  a_2 \Delta_U-b_1 \Delta_S - C_{1} \Delta_T) \hspace{2mm} , \hspace{2mm} \, c_{0}  \rightarrow  \lambda_T (c_0+C_{1} \Delta_U - 2 c_{3}' \Delta_T-d_0 \Delta_S) \\[2mm]
b_{0}  \rightarrow  \lambda_S (b_0-3 b_1 \Delta_U+3 d_0 \Delta_T) \hspace{2mm} \,,\, \hspace{2mm}  a_{2}  \rightarrow  \lambda_U^2 a_2  \hspace{2mm} \,,\, \hspace{2mm}   b_{1}  \rightarrow  \lambda_S \lambda_U b_1 
\hspace{2mm} \,,\, \hspace{2mm}  C_{1}  \rightarrow  \lambda_T \lambda_U C_{1}\\[2mm]
c_{3}'  \rightarrow  \lambda_{T}^2  c_{3}'  \hspace{3mm} , \hspace{3mm}   d_{0}  \rightarrow  \lambda_{S}  \lambda_{T} d_{0}  \,\, \  ,
\end{array}
\eeq
with $C_{1}=(2 c_1 - \tilde{c}_{1})$. Therefore two different M-theory flux vacua among those found in the previous section can be viewed as critical points of the same supergravity model -- one of them this time moved outside the origin -- if their corresponding M-theory flux backgrounds are connected via the transformations (\ref{Flux_redefinition}). In other words, if they belong to the same \mbox{$[\textrm{SU}(1,1)/\textrm{U}(1)]^3$-duality} orbit of flux backgrounds. 

It is then easy to check\footnote{Notice that the overall scaling parameter $\lambda$ might be different (and generically will be) in two flux backgrounds related by the transformations (\ref{Flux_redefinition}).} that the $17$ different vacua we found in the previous section belong to uniquely four duality orbits of flux backgrounds. More concretely they group as
\beq
\begin{array}{lllllllllllll}
\textrm{Model 1}& : & \textrm {vac 1} & \leftrightarrow & \textrm {vac 3} & \leftrightarrow &\textrm {vac 5}  \\[2mm]
\textrm{Model 2} & : & \textrm {vac 2} & \leftrightarrow &\textrm {vac 4} & \leftrightarrow &\textrm {vac 6} & \leftrightarrow &\textrm {vac 7} & \leftrightarrow & \textrm {vac 8} \\[2mm]
\textrm{Model 3} & : & \textrm {vac 9} & \leftrightarrow &\textrm {vac 12} & \leftrightarrow &\textrm {vac 13} & \leftrightarrow &\textrm {vac 14} & \leftrightarrow & \textrm {vac 15} & \leftrightarrow & \textrm {vac 17} \\[2mm]
\textrm{Model 4} & : & \textrm {vac 10} & \leftrightarrow &\textrm {vac 11} & \leftrightarrow &\textrm {vac 16}
\end{array} \nonumber
\eeq
where we have identified one duality orbit with one supergravity model. The $\,\mathcal{N}=4\,$ gauging underlying each of the four inequivalent  STU-models can be computed by looking at any of the orbit representatives. We decide to select the first representative in each of the models. They are given by
\beq
\begin{array}{lllllllll}
\textrm{Model 1}& : & a_{0} = -3 \, \lambda & , &  a_{1} = b_{0} = c_{0} = 0 & , &  b_{1} = - d_{0} =  \lambda & \ , \\[2mm]
\textrm{Model 2}& : & a_{0} = \phantom{-}3 \, \lambda & , &  a_{1} = b_{0} = c_{0} = 0 & , &  b_{1} = - d_{0} =  \lambda & \ , \\[2mm]
\textrm{Model 3}& : & a_{0} = - 3 \, \lambda & , &  a_{1} = b_{0} = c_{0} = 0 & , &  b_{1} = - d_{0} =  \lambda & , &  a_{2} = - c_{3}' =  \frac{\lambda}{2}   \ , \\[2mm]
\textrm{Model 4}& : & a_{0} = \phantom{-} 3 \, \lambda & , &  a_{1} = b_{0} = c_{0} = 0 & , &  b_{1} = - d_{0} =  \lambda & , &  a_{2} = - c_{3}' =  \frac{\lambda}{2}   \ , \nonumber
\end{array}
\eeq
together with (\ref{fluxes_common_onlyKK6}) for Model 1 and 2 as well as (\ref{fluxes_common_bothKK6}) for Model 3 and 4. A detailed description of these M-theory backgrounds is collected in the appendices. Remarkably they require $a_{1} = b_{0} = c_{0} = 0$ implying a vanishing flux $\,G_{(4)}=0\,$ (see Table~\ref{Table:M-Theory_fluxes}) as well as a non-vanishing $\,G_{(7)}\neq 0\,$. The result is then an M-theory superpotential (\ref{WMtheory}) of the form
\beq
\label{WMtheory_no_linear}
W_{\textrm{M-theory}} = \frac{1}{4} \int_{X_{7}} G_{(7)} + \frac{1}{8} \int_{X_{7}} (A_{(3)} + i \Phi_{(3)}) \wedge  d (A_{(3)} + i \Phi_{(3)}) \ ,
\eeq
uniquely induced by metric $\omega$ and $\,G_{(7)}\,$ background fluxes. The underlying $\mathcal{N}=4$ gauging turns out to have a very simple algebra structure as we will investigate now.

The four inequivalent STU-models can be simultaneously explored by considering the gauge algebra \mbox{$\,G_{0} \subset \textrm{SL}(2) \times \textrm{SO}(6,6)\,$}  induced by the set of seven fluxes
\beq
\begin{array}{lllllllll}
a_{0}=-{f_{+}}^{abc} & , & a_{2}= -{f_{+}}^{ajk} & , &  \tilde{c}_{1}=f_{+\phantom{bc} a}^{\phantom{+}bc} & , & c_{1}= f_{+\phantom{aj} k}^{\phantom{+}aj} = f_{+\phantom{ka} j}^{\phantom{+}ka} & , & c_{3}'= {f_{+jk}}^{a}  \\[2mm]
b_{1}= {f_{-}}^{ibc} & , & d_{0}= f_{-\phantom{bc} \,i}^{\phantom{-}bc} & ,
\end{array}
\eeq
where the upper line contains electric fluxes ($\alpha=+$) and the lower line magnetic ones ($\alpha=-$). In order to analyse the structure of the gauge brackets in (\ref{BracketsN=4}), we use again the splitting of the SO(6,6) light-cone index $\,{_M=(\,_a\,,\,_i\,,\,^a\,,\,^i\,)}\,$ in the generators $\,T_{\alpha M}$ 
\beq
\begin{array}{lllllllll}
T_{+a} \equiv Z_{a} & , & T_{+i} \equiv Z_{i} & , & {T_{+}}^{a} \equiv X^{a} & , & {T_{+}}^{i} \equiv X^{i} & , \\[2mm]
T_{-a} \equiv \bar{Z}_{a} & , & T_{-i} \equiv \bar{Z}_{i} & , & {T_{-}}^{a} \equiv \bar{X}^{a} & , & {T_{-}}^{i} \equiv \bar{X}^{i} & .
\end{array}
\eeq
Using this decomposition, the antisymmetry of the brackets in (\ref{BracketsN=4}) determines the magnetic generators in terms of the electric ones so that only an independent twelve-dimensional algebra is gauged. One obtains
\beq
\begin{array}{lllllllll}
\bar{Z}_{a} = 0 & , & \bar{Z}_{i} = \dfrac{c_{1} d_{0}- b_{1} c_{3}'}{c_{1}^2 +a_{2} c_{3}'} \, Z_{a} & , & \bar{X}^{a} = \dfrac{b_{1}}{\tilde{c}_{1}} \, Z_{i} + \dfrac{d_{0}}{\tilde{c}_{1}} \, X^{i} & , & \bar{X}^{i} = \dfrac{ b_{1} c_{1} + a_{2} d_{0}}{c_{1}^2 +a_{2} c_{3}'} \, Z_{a} \,\ .
\end{array}
\eeq
According to the Levi decomposition of finite dimensional real lie algebras, we find that $G_{0} = G_{\textrm{semi}} \ltimes G_{\textrm{solv}}$ with a $3$-dimensional semisimple piece and a $9$-dimensional solvable piece. The generators $\{ Z_{i} , X^{i} \, ; \, Z_{a} \}$ span the solvable (actually nilpotent) ideal with non-vanishing brackets
\beq
[Z_{i} , Z_{j} ] = c_{3}' \, Z_{c}
\hspace{5mm} , \hspace{5mm}
[Z_{i} , X^{j} ] = c_{1} \, Z_{c}
\hspace{5mm} , \hspace{5mm}
[X^{i} , X^{j} ] = - a_{2} \, Z_{c} \ ,
\eeq
which can be understood as $G_{\textrm{solv}} = U(1)^{6} \ltimes U(1)^{3}$. The mixed brackets between the solvable and the semisimple pieces read
\beq
[Z_{a} , X^{b} ] = \tilde{c}_{1} \, Z_{c}
\hspace{5mm} , \hspace{5mm}
[Z_{i} , X^{b} ] = c_{1} \, Z_{k} + c_{3}' \, X^{k}
\hspace{5mm} , \hspace{5mm}
[X^{i} , X^{b} ] = - a_{2} \, Z_{k} + c_{1} \, X^{k} \ ,
\eeq
whereas the non-vanishing commutators between generators $X^{a}$ in the semisimple piece are given by
\beq
[X^{a} , X^{b} ] = \tilde{c}_{1} \, X^{c} - a_{0} \, Z_{c} \ .
\eeq
The full $12$-dimensional $\mathcal{N}=4$ gauging thus corresponds to a gauge group
\beq
G_{0}= \textrm{SO(3)} \ltimes \textrm{Nil}_{9}(2) \ ,
\eeq
with $\textrm{Nil}_{9}(2)$ being a $9$-dimensional $\textrm{U}(1)^{6} \ltimes \textrm{U}(1)^{3}$ nilpotent ideal of order two (three steps) with lower central series
\beq
\{ Z_{i} \, , \,  X^{i} \, , \, Z_{a} \} \supset \{  Z_{a} \} \supset 0 \ .
\eeq

The different values of the fluxes in the four disconnected STU-models only determine how the semisimple products are specifically realised but do not modify the identification of the full group as $G_{0}= \textrm{SO(3)} \ltimes \textrm{Nil}_{9}(2)$. This gauge group (with a different realisation in terms of brackets) has also appeared in twisted reductions of \textit{massive} type IIA strings \cite{Dall'Agata:2009gv,Dibitetto:2011gm}. However, as we already emphasised, those \textit{massive} type IIA backgrounds cannot be obtained from our M-theory reductions due to the lack of the Romans mass parameter\footnote{The Romans mass parameter $a_{3}$ generates the only cubic coupling $\,-a_{3}\, U^{3}\,$ in the IIA superpotential (\ref{Superpotential_Flux_IIA_ISO}). This term cannot be removed by applying the STU duality transformations in (\ref{non-compact_transf_ISO2}), hence forcing the \textit{massive} IIA backgrounds to lie in a different duality orbit of STU-models than the M-theory backgrounds.}. 

\subsection{Overview of M-theory backgrounds with monopoles}
\label{overview}

Let us summarise the set of M-theory flux vacua we have obtained when including monopoles in the background and also discuss their main features. Turning on KK6 (KKO6)  monopoles compatible with $\mathcal{N}=4$ supersymmetry of the effective action, we have found three different situations:

\begin{itemize}

\item[$1)$] In the first situation the M-theory background is compatible with having only KKO6-planes admitting an interpretation in terms of O6$_{\parallel}$-planes in a type IIA orientifold incarnation of the effective STU-model. The effective flux model is a no-scale supergravity and corresponds to the ``\textbf{vac 0}" solution in Table~\ref{Table:Landscape}. This scenario was discussed in detail in section~\ref{sec:only D6}.

\item[$2)$] The second situation involves M-theory backgrounds compatible with having only KKO6$_{\parallel}$-planes, hence lacking a type IIA interpretation in terms of an ordinary STU orientifold model.  There are $8$ different backgrounds lying inside two different duality orbits of theories (inequivalent superpotentials) after making use of duality transformations in the effective action. Full moduli stabilisation can be achieved in this type of backgrounds producing supersymmetric ($\mathcal{N}=1,3$) as well as non-supersymmetric and stable AdS$_{4}$ vacua. An analysis of the $G_{2}$-structure underlying these solutions reveals the existence of one supersymmetric ($\mathcal{N}=1$) and two non-supersymmetric backgrounds with weak $G_{2}$ holonomy. The rest correspond to $G_{2}$-structures where torsion classes both in the $\textbf{1}$ ($\widetilde{W}_{1}$) and the $\textbf{27}$ ($\widetilde{W}_{27}$) are activated. In all the M-theory backgrounds within this category the $\mathcal{N}=4$ gauging  $G_{0}=\textrm{SO(3)} \ltimes \textrm{Nil}_{9}(2)$ is broken to a $\,G_{\textrm{res}}=\textrm{SO}(3)\,$ subgroup at the vacuum. In addition, some unstable solutions also exist. The results are summarised in the second block of Table~\ref{Table:Landscape} encompassing solutions from $\textbf{vac 1}$ to $\textbf{vac 8}$.

\item[$3)$] The third and last situation corresponds to M-theory backgrounds compatible with having simultaneously O6$_{\parallel}$-planes as well as KKO6$_{\parallel}$-planes. Due to the necessity of the latter, no type IIA interpretation in terms of ordinary STU orientifold models is possible either.  There are $9$ different M-theory backgrounds -- from \mbox{\textbf{vac 9}} to \textbf{vac 17} -- lying inside two duality orbits of inequivalent theories also with an $\mathcal{N}=4$ gauging $G_{0}=\textrm{SO(3)} \ltimes \textrm{Nil}_{9}(2)$, although it has a slightly different realisation in terms of gauge brackets (see discussion in sec.~\ref{sec:N=4_gauging}).  One remarkable consequence of having the two types of sources is that the residual symmetry group gets enhanced to $\,G_{\textrm{res}}=\textrm{SO}(3) \ltimes U(1)^3\,$ at the supersymmetric ($\mathcal{N}=1,3$) solutions as well as at the three non-supersymmetric solutions \textbf{vac 10} (stable), \textbf{vac 12} (stable) and \textbf{vac 13} (unstable) in Table~\ref{Table:Landscape}. The rest of the solutions in this category turn out to be non-supersymmetric, unstable and preserve $\,G_{\textrm{res}}=\textrm{SO}(3)$. Concerning the $G_{2}$-structure of these backgrounds, the situation is similar to the previous case: there is one supersymmetric ($\mathcal{N}=1$) and two non-supersymmetric M-theory backgrounds with weak $G_{2}$ holonomy. The rest of the solutions activate $\widetilde{W}_{1}$ as well as $\widetilde{W}_{27}$. These results are collected in the third block of Table~\ref{Table:Landscape}.

\end{itemize}

The above set of M-theory backgrounds has been obtained after relaxing the ordinary Scherk-Schwarz conditions (\ref{SS_Constraints}) in a way compatible with $\mathcal{N}=4$ supersymmetry in the effective action. One could completely forget about the entire set of conditions in (\ref{SS_Constraints}) if permitting all the types of sources in Table~\ref{Table:KK6} to be present in the background. The resulting theory would then just preserve $\mathcal{N}=1$ supersymmetry. In this case new solutions might (and generically will) appear involving more complex configurations of sources in higher dimensions and consequently more elaborated flux backgrounds and superpotentials in four dimensions. In addition, due to the rich structure of moduli powers in the scalar potential (see section~\ref{sec:scalings}), one might hope for the existence of de Sitter solutions in these M-theory scenarios. However a no-go theorem forbidding the existence of such solutions can be derived along the lines of ref.~\cite{Hertzberg:2007wc} using the M-theory universal moduli \cite{Dibitetto&Danielsson}. Even if charting the landscape of such unrestricted M-theory configurations could be too ambitious, one could still restrict the scan to solutions preserving weak $G_{2}$ holonomy or some other restricted structure of torsion classes. For the sake of simplicity, we have just restricted ourselves in this work to M-theory backgrounds preserving $\,\mathcal{N}=4\,$ supersymmetry in four dimensions.

\section{Summary and final remarks}
\label{sec:conclusions}

In this paper we have investigated M-theory reductions on $G_{2}$-manifolds in the presence of gauge and metric background fluxes as well as KK6 (KKO6) sources from a purely four-dimensional point of view. We have done it in terms of the $\,\mathcal{N}=1\,$ effective STU-models describing truncations of $\,\mathcal{N}=4\,$ gauged supergravity on the basis of the embedding tensor formalism \cite{Schon:2006kz,Dibitetto:2011eu}. 

In the first part of the paper we investigated the interplay between the conditions (\ref{SS_Constraints}) required by an ordinary Scherk-Schwarz reduction and the consistency relations (\ref{QC_N=4}) and (\ref{QC_N=8extra}) imposed by $\,\mathcal{N}=4\,$ and $\,\mathcal{N}=8\,$ supersymmetry on the corresponding gauged supergravity. The outcome was that while demanding $\,\mathcal{N}=8\,$ supersymmetry in the effective action amounts to imposing the entire set of conditions in  (\ref{SS_Constraints}),  requiring only $\,\mathcal{N}=4\,$ allows for a relaxation of some of the Scherk-Schwarz conditions. More concretely, the conditions $iv)$ and $vii)$ in the list of (\ref{ww_constraint_iso}). The non-vanishing of these two conditions was respectively linked to the presence of  D6$_\parallel$ (O6$_\parallel$) and KK6$_\parallel$ (KKO6$_\parallel$) sources in the background (see Table~\ref{Table:KK6}) and also to the activation of the genuine M-theory fluxes $\,(c_{3}'\,,\,d_{0})\,$ in the superpotential (\ref{Superpotential_comparison_ISO_Intro}) which had no counterpart in the type IIA orientifold constructions of refs~\cite{Derendinger:2004jn,Villadoro:2005cu,Dall'Agata:2009gv,Dibitetto:2011gm}. The novel KK6$_\parallel$ (KKO6$_\parallel$) sources  were found to induce new universal moduli powers in the scalar potential (\ref{V_M-theory}) thus opening new possibilities for moduli stabilisation in M-theory flux models.

In the second part of the paper we performed a systematic and exhaustive study of \mbox{M-theory} flux vacua by combining the use of dualities in the STU-models with algebraic geometry tools available in the computer algebra system \textsc{singular}. We proved that full moduli stabilisation can be achieved in $\,\mathcal{N}=4\,$ flux models coming from M-theory provided KKO6-planes are included as background sources (see Table~\ref{Table:Landscape} and section~\ref{overview} for a summary of the results). The underlying $\,\mathcal{N}=4\,$ gauging is unique and identified with $\,G_{0}= \textrm{SO(3)} \ltimes \textrm{Nil}_{9}(2)$. Moreover, we also showed that these models correspond to genuine M-theory backgrounds  which do not admit an interpretation in terms of regular type IIA orientifold constructions. In the latter, moduli stabilisation seems to demand a non-vanishing Romans mass parameter \cite{Romans:1985tz} and therefore a deformation already in higher dimensions \cite{Dall'Agata:2009gv,Dibitetto:2011gm}. This deformation parameter does not appear from M-theory upon ordinary dimensional reduction so, in the M-theory backgrounds we have found here, full moduli stabilisation is achieved from a massless theory in higher dimension. Moreover a background flux for the $G_{(7)}$ form -- $a_{0}$ parameter in (\ref{Superpotential_Flux_MTheory_ISO}) -- seems to be mandatory in this case, thus playing a similar role as the Romans mass for moduli stabilisation in a type IIA context but having a neat ``field strength" interpretation in higher dimensions as the dual of a purely external $\,G_{(4)}\,$ flux \cite{Freund:1980xh,DallAgata:2005fm}.  

Finally we want to stress once more the four-dimensional, bottom-up approach adopted in this paper which justifies to adopt the ET formalism to analyse the∫ effect of the particular M-theory fluxes without a type IIA interpretation, namely, M-theory fluxes becoming \textit{non-geometric} in the IIA picture. While focusing on four-dimensional solutions has interest for obvious reason, one may feel concerned with 
the actual existence of a ten-dimensional type IIA interpretation of the flux vacua we found here. The ultimate connection should be between strings/M-theory and four-dimensional physics, and insisting on an intermediate ten-dimensional field theory step could appear somewhat artificial\footnote{Four-dimensional gauged supergravities have been shown to capture the dynamics of asymmetric orbifold constructions for which a ten-dimensional ``geometry" does not even exist due to the difference between the left $X_{L}$ and right $X_{R}$ sectors. Further interesting connections to non-geometric flux backgrounds have also been established in refs~\cite{Dabholkar:2002sy,Dabholkar:2005ve,Condeescu:2013yma}.}. On the other hand, because of the 16 supercharges they preserve, it would be interesting to explore potential realisations of the gauged supergravities studied here as eleven-dimensional $1/2$-BPS backgrounds. We find the reach structure of M-theory flux vacua presented in this work an additional motivation for pursuing this goal. We hope to come back to this in the future.

%
%

\section*{Acknowledgments}

We want to thank Ulf Danielsson, Giuseppe Dibitetto and Diederik Roest for interesting discussions and specially Giuseppe Dibitetto for the careful reading of an early draft of this paper. The work of the authors is supported by the Swiss National Science Foundation.

%
%

\appendix


\section{Backgrounds with only KK6$_{\parallel}$ (KKO6$_{\parallel}$)}
\label{app:only_KK6_parallel}

In this first appendix we present the detailed analysis of the eight M-theory backgrounds in section~\ref{sec:KK6_parallel only} which include only KK6 (KKO6) $\rightarrow$ KK6$_{\parallel}$ (KKO6$_{\parallel}$) sources. They all are compatible with the flux condition (\ref{fluxes_common_onlyKK6}). We present the associated background fluxes, M-theory superpotential, vacuum energy $V_{0}$, normalised\footnote{We normalise the masses with respect to the AdS$_{4}$ radius $\,L=\sqrt{-3/V_{0}}\,$ so that perturbative stability requires tachyons to satisfy the Breitenlohner-Freedman (BF) \cite{Breitenlohner:1982bm} bound $\,m^2 L^2 \geq -9/4\,$.} mass spectra for scalars and vectors and preserved supersymmetry at each of the M-theory flux vacua.

\subsubsection*{Vacuum $1$}

This solution corresponds to the flux configuration
\beq
\label{Fluxes_vacuum1}
a_{0} = -3 \, \lambda \,\,\,\, , \,\,\,\, a_{1} = b_{0} = c_{0} = 0 \,\,\,\, , \,\,\,\,  b_{1} = - d_{0} =  \lambda \ ,
\eeq
which implies a superpotential (\ref{Superpotential_Flux_MTheory_ISO}) of the form
\beq
\label{Superpotential_vacuum1}
W^{\textrm{(iso)}}_{\textrm{M-Theory}} = 3 \, \lambda \, (S \, T + S \,  U + T \, U - 1)  \ .
\eeq
The value of the potential evaluated at this critical points is $V_{0}=-3 \lambda^2/8$ corresponding to an AdS$_{4}$ solution. The scalar masses turn out to be
\beq
\label{vacuum1_scalars}
m^2 L^2 = 18 \,\, (\times 1) \,\,\, , \,\,\, 10 \,\, (\times 6) \,\,\, , \,\,\, 4 \,\, (\times 7) \,\,\, , \,\,\, -2 \,\, (\times 5) \,\,\, , \,\,\, 0 \,\, (\times 19) \ ,
\eeq
not displaying instabilities but displaying ten flat directions not associated to Goldstone bosons. The vector masses read
\beq
\label{vacuum1_vectors}
m^2 L^2 = 12 \,\, (\times 3) \,\,\, , \,\,\, 6 \,\, (\times 3) \,\,\, , \,\,\, 2 \,\, (\times 3) \,\,\, , \,\,\, 0 \,\, (\times 3) \ ,
\eeq
where the three massless vectors are associated to the SO(3) residual symmetry preserved by the STU models. The computation of the gravitini mass matrix (\ref{gravitini_mass_matrix}) reveals $\mathcal{N}=3$ preserved supersymmetry.

\subsubsection*{Vacuum $2$}

This solution simply changes the sign of $a_{0}$ compared to (\ref{Fluxes_vacuum1})
\beq
\label{Fluxes_vacuum2}
a_{0} = 3 \, \lambda \,\,\,\, , \,\,\,\, a_{1} =  b_{0} = c_{0} = 0 \,\,\,\, , \,\,\,\,  b_{1} = -d_{0} =  \lambda \ ,
\eeq
so the superpotential now reads
\beq
\label{Superpotential_vacuum2}
W^{\textrm{(iso)}}_{\textrm{M-Theory}} = 3 \, \lambda \, (S \, T + S \,  U + T \, U + 1)  \ .
\eeq
At this critical point the value of the energy is also $V_{0}=- 3 \lambda^2/8$. The computation of the scalar masses
\beq
\label{vacuum2_scalars}
m^2 L^2 = 18 \,\, (\times 2) \,\,\, , \,\,\, 10 \,\, (\times 5) \,\,\, , \,\,\, 4 \,\, (\times 6) \,\,\, , \,\,\, -2 \,\, (\times 7) \,\,\, , \,\,\, 0 \,\, (\times 18) \ ,
\eeq
gives different multiplicities compared to (\ref{vacuum1_scalars}) and nine flat directions. The vector masses are given by
\beq
\label{vacuum2_vectors}
m^2 L^2 = 12 \,\, (\times 3) \,\,\, , \,\,\, 6 \,\, (\times 3) \,\,\, , \,\,\, 2 \,\, (\times 3) \,\,\, , \,\,\, 0 \,\, (\times 3) \ ,
\eeq
coinciding with those in (\ref{vacuum1_vectors}). However, the computation of the gravitini mass matrix (\ref{gravitini_mass_matrix}) shows that supersymmetry is completely broken at this solution.

\subsubsection*{Vacuum $3$}

This solution is compatible with the family of flux parameters
\beq
\label{Fluxes_vacuum3}
a_{0} = -\frac{9}{5} \, \lambda \,\,\,\, , \,\,\,\, a_{1} =  b_{0} = c_{0}  = 0 \,\,\,\, , \,\,\,\,  b_{1} = - d_{0} =  \frac{1}{5} \, \lambda \ ,
\eeq
so this time the superpotential is given by
\beq
\label{Superpotential_vacuum3}
W^{\textrm{(iso)}}_{\textrm{M-Theory}} = \frac{3}{5} \, \lambda \left( S T+S U+5\, T U - 3 \right)  \ .
\eeq
The value of the vacuum energy at this solution is $V_{0}=- 27 \lambda^2/200$. Computation of the scalar masses we find
\beq
\label{vacuum3_scalars}
\begin{array}{ccll}
m^2 L^2 &=& \frac{190}{9} \,\, (\times 5) \,\,\, , \,\,\, 18 \,\, (\times 1) \,\,\, , \,\,\, \frac{112}{9}  \,\, (\times 5) \,\,\, , \,\,\, \frac{70}{9}  \,\, (\times 1) \,\,\, , \,\,\, \frac{52}{9}  \,\, (\times 5) \,\,\, , \,\,\, \frac{32}{9}  \,\, (\times 3) & ,\\[2mm]
& & -\frac{20}{9} \,\, (\times 1) \,\,\, , \,\,\, -2 \,\, (\times 1) \,\,\, , \,\,\, \frac{10}{9}  \,\, (\times 6) \,\,\, , \,\,\, -\frac{8}{9}  \,\, (\times 1) \,\,\, , \,\,\, 0  \,\, (\times 9) & ,
\end{array}
\eeq
so there are neither instabilities (tachyons satisfy the BF bound) nor flat directions at this vacuum (the nine massless scalars correspond to Goldstone bosons). For the vector masses we obtain
\beq
\label{vacuum3_vectors}
m^2 L^2 = 12 \,\, (\times 3) \,\,\, , \,\,\, 6 \,\, (\times 3) \,\,\, , \,\,\, \tfrac{50}{9} \,\, (\times 3) \,\,\, , \,\,\, 0 \,\, (\times 3) \ ,
\eeq
displaying three massless vectors associated to the SO(3) residual symmetry. Supersymmetry is completely broken at this solution.

\subsubsection*{Vacuum $4$}

This solution is very similar to (\ref{Fluxes_vacuum3}). Again there is a sign flip for the $a_{0}$ flux parameter. The background fluxes is given by
\beq
\label{Fluxes_vacuum4}
a_{0} = \frac{9}{5} \, \lambda \,\,\,\, , \,\,\,\, a_{1} = b_{0} = c_{0}  =  0 \,\,\,\, , \,\,\,\,  b_{1} = - d_{0} =  \frac{1}{5} \, \lambda \ ,
\eeq
so the superpotential reads
\beq
\label{Superpotential_vacuum4}
W^{\textrm{(iso)}}_{\textrm{M-Theory}} = \frac{3}{5} \, \lambda \left( S T+S U+5\, T U + 3 \right)  \ ,
\eeq
and the vacuum energy at this solution is also $V_{0}=- 27 \lambda^2/200$. The mass spectrum for the scalars turns out to be
\beq
\label{vacuum4_scalars}
\begin{array}{ccll}
m^2 L^2 &=& \frac{190}{9} \,\, (\times 5) \,\,\, , \,\,\, 18 \,\, (\times 1) \,\,\, , \,\,\,  \frac{112}{9}  \,\, (\times 5) \,\,\, , \,\,\, 10  \,\, (\times 1) \,\,\, , \,\,\, \frac{52}{9}  \,\, (\times 5) \,\,\, , \,\,\, \frac{32}{9}  \,\, (\times 3) & ,\\[2mm]
& & -\frac{20}{9} \,\, (\times 2) \,\,\, , \,\,\,  \frac{10}{9}  \,\, (\times 5) \,\,\, , \,\,\, -\frac{8}{9}  \,\, (\times 2) \,\,\, , \,\,\, 0  \,\, (\times 9) & ,
\end{array}
\eeq
again featuring neither instabilities nor flat directions. The spectrum of vector masses is
\beq
\label{vacuum4_vectors}
m^2 L^2 = 12 \,\, (\times 3) \,\,\, , \,\,\, 6 \,\, (\times 3) \,\,\, , \,\,\, \tfrac{50}{9} \,\, (\times 3) \,\,\, , \,\,\, 0 \,\, (\times 3) \ ,
\eeq
and coincides with that of (\ref{vacuum3_vectors}). Substituting the values of the fluxes into the gravitini mass matrix (\ref{gravitini_mass_matrix}), this solution happens to preserve $\,\mathcal{N}=1\,$ supersymmetry.

\subsubsection*{Vacuum $5$}

This solution corresponds to background fluxes of the form
\beq
\label{Fluxes_vacuum5}
a_{0} = -\frac{6}{5} \, \lambda \,\,\,\, , \,\,\,\, a_{1} = b_{0} = -c_{0} = \frac{3}{5} \, \lambda \,\,\,\, , \,\,\,\,  b_{1} = - d_{0} =  \frac{1}{5} \, \lambda \ ,
\eeq
and induces the flux superpotential
\beq
\label{Superpotential_vacuum5}
W^{\textrm{(iso)}}_{\textrm{M-Theory}} = \frac{3}{5} \, \lambda \left( -S -3 T -3\, U + S T+S U+ 5\, T U - 2 \right)  \ .
\eeq
It produces a vacuum energy $V_{0}=- 9 \lambda^2/80$. We find the following mass spectrum for the scalars
\beq
\label{vacuum5_scalars}
\begin{array}{ccll}
m^2 L^2 &=& \frac{76}{3} \,\, (\times 5) \,\,\, , \,\,\, 18 \,\, (\times 1) \,\,\, , \,\,\,  \frac{64}{3}  \,\, (\times 5) \,\,\, , \,\,\, \frac{28}{3}  \,\, (\times 1) \,\,\, , \,\,\, \frac{20}{3}  \,\, (\times 3) \,\,\, , \,\,\, 6  \,\, (\times 1) & ,\\[2mm]
& & \frac{16}{3} \,\, (\times 5) \,\,\, , \,\,\,  \frac{4}{3}  \,\, (\times 6) \,\,\, , \,\,\, -\frac{2}{3}  \,\, (\times 2) \,\,\, , \,\,\, 0  \,\, (\times 9) & ,
\end{array}
\eeq
which does not present neither instabilities nor flat directions. The vector mass spectrum is given by
\beq
\label{vacuum5_vectors}
m^2 L^2 = 12 \,\, (\times 6) \,\,\, , \,\,\, \tfrac{20}{3}  \,\, (\times 3)  \,\,\, , \,\,\, 0 \,\, (\times 3) \ ,
\eeq
containing the three SO(3) massless vectors. The gravitini mass matrix shows that supersymmetry is completely broken at this solution.

\subsubsection*{Vacuum $6$}

This solution is generated by a flux background given by
\beq
\label{Fluxes_vacuum6}
a_{0} = \lambda  \,\,\,,\,\,\,  a_{1} =c_{0}=  -\tfrac{1}{2} \, \sqrt{ \tfrac{1}{6} \,(31+3\,\sqrt{57}) }\, \lambda   \,\,\,,\,\,\,    b_{0} = 0   \,\,\,,\,\,\,   b_{1} = -d_{0} = \tfrac{1}{12} \, (9+\sqrt{57}) \, \lambda \ ,
\eeq
and the associated superpotential lacks the linear term on the dilaton modulus $S$. The vacuum energy at this solution is $V_{0}=- \tfrac{1}{64} \, (13 + \sqrt{57})\, \lambda^2$. Computing the masses for the scalars we obtain
\beq
\label{vacuum6_scalars}
\begin{array}{ccll}
m^2 L^2 &=& 27.976 \,\, (\times 1) \,\,\, , \,\,\, 18 \,\, (\times 1) \,\,\, , \,\,\,  16.921  \,\, (\times 5) \,\,\, , \,\,\, 13.771  \,\, (\times 1)  & ,\\[2mm]
& & 10.107 \,\, (\times 1) \,\,\, , \,\,\, 10.083 \,\, (\times 5)  \,\,\, , \,\,\,  9.221 \,\, (\times 1) \,\,\, , \,\,\, 5.557  \,\, (\times 3)  & , \\[2mm]
& & -3.961  \,\, (\times 5) \,\,\, , \,\,\, -2.648  \,\, (\times 1) \,\,\, , \,\,\, 2.086  \,\, (\times 5) \,\,\, , \,\,\, 0  \,\, (\times 9) & .
\end{array}
\eeq
This time there are tachyon masses violating the BF bound, thus rendering this solution unstable. Still there are no flat directions as the nine massless scalars correspond to Goldstone modes. The computation of the vector masses yields
\beq
\label{vacuum6_vectors}
m^2 L^2 = 12 \,\, (\times 3) \,\,\, , \,\,\, \tfrac{1}{7}\, (97 - \sqrt{57})  \,\, (\times 3)   \,\,\, , \,\,\, \tfrac{1}{7}\, (71 + \sqrt{57})  \,\, (\times 3)   \,\,\, , \,\,\, 0 \,\, (\times 3) \ ,
\eeq
containing the three massless vectors of the SO(3) residual symmetry. This is a non-supersymmetric and unstable solution.

\subsubsection*{Vacuum $7$}

This solution is induced by a family of fluxes of the form
\beq
\label{Fluxes_vacuum7}
a_{0} = \lambda  \,\,\,,\,\,\,  a_{1} =c_{0}=  -\tfrac{1}{2} \, \sqrt{ \tfrac{1}{6} \,(31-3\,\sqrt{57}) }\, \lambda   \,\,\,,\,\,\,    b_{0} = 0   \,\,\,,\,\,\,   b_{1} = -d_{0} = \tfrac{1}{12} \, (9-\sqrt{57}) \, \lambda   \ ,
\eeq
then being very similar to the previous solution (\ref{Fluxes_vacuum6}) and lacking also the linear term on $S$. The vacuum energy is nevertheless different $V_{0}=- \tfrac{1}{64} \, (13 - \sqrt{57})\, \lambda^2$. The spectrum of scalar masses is given by
\beq
\label{vacuum7_scalars}
\begin{array}{ccll}
m^2 L^2 &=& 31.857 \,\, (\times 5) \,\,\, , \,\,\, 18 \,\, (\times 1) \,\,\, , \,\,\,  30.364  \,\, (\times 5) \,\,\, , \,\,\, 12.686  \,\, (\times 1)  & ,\\[2mm]
& & 9.871 \,\, (\times 3) \,\,\, , \,\,\,  7.064 \,\, (\times 1)  \,\,\, , \,\,\,  6.564 \,\, (\times 5) \,\,\, , \,\,\, 3.799  \,\, (\times 5)  & , \\[2mm]
& & -3.486  \,\, (\times 1) \,\,\, , \,\,\, -0.678  \,\, (\times 1) \,\,\, , \,\,\, -0.300  \,\, (\times 1) \,\,\, , \,\,\, 0  \,\, (\times 9) & ,
\end{array}
\eeq
and features one tachyon mass violating the BF bound. The solution is then unstable and does not contain flat directions. The set of vector masses reads
\beq
\label{vacuum7_vectors}
m^2 L^2 = 12 \,\, (\times 3) \,\,\, , \,\,\, \tfrac{1}{7}\, (97 + \sqrt{57})  \,\, (\times 3)   \,\,\, , \,\,\, \tfrac{1}{7}\, (71 - \sqrt{57})  \,\, (\times 3)   \,\,\, , \,\,\, 0 \,\, (\times 3) \ ,
\eeq
being then similar to (\ref{vacuum6_vectors}) and, as usual, showing the three massless vectors of the SO(3) symmetry. Computing the gravitini mass matrix (\ref{gravitini_mass_matrix}) one finds that this solution is non-supersymmetric.

\subsubsection*{Vacuum $8$}

The last solution of this section is associated to the most complex family of background fluxes
\beq
\label{Fluxes_vacuum8}
\begin{array}{llllll}
a_{0} = \tfrac{5}{4} \, \lambda & , & a_{1} = \tfrac{1}{24} (\sqrt{15}-3\,\sqrt{35}) \, \lambda & , &  b_{0} = -\tfrac{\sqrt{15}}{4} \, \lambda & , \\[2mm]
b_{1} = \tfrac{1}{24}(9-\sqrt{21}) \, \lambda  &  , &  c_{0} =  -\frac{1}{4} \sqrt{ \frac{5}{6} \,(11+\sqrt{21}) } \, \lambda  & , &   d_{0} =  -\tfrac{1}{24} \, (9+\sqrt{21}) \, \lambda &   .
\end{array}
\eeq
The flux-induced superpotential consists of the same moduli couplings as (\ref{Superpotential_vacuum5}) this time specified by (\ref{Fluxes_vacuum6}), and produces a vacuum energy $V_{0}=- 5 \lambda^2/32$. The scalar mass spectrum is given by\footnote{For the sake of clarity we display the numerical value of irrational numbers.}
\beq
\label{vacuum8_scalars}
\begin{array}{ccll}
m^2 L^2 &=& 20.955 \,\, (\times 5) \,\,\, , \,\,\, 18 \,\, (\times 1) \,\,\, , \,\,\,  14.341  \,\, (\times 5) \,\,\, , \,\,\, 14.181  \,\, (\times 1)  & ,\\[2mm]
& & 3.986 \,\, (\times 5) \,\,\, , \,\,\, 2.539 \,\, (\times 1)  \,\,\, , \,\,\, -1.150  \,\, (\times 1) \,\,\, , \,\,\, -1.120  \,\, (\times 1)  & , \\[2mm]
& & \frac{24}{5}  \,\, (\times 3) \,\,\, , \,\,\, -0.882  \,\, (\times 5) \,\,\, , \,\,\, 8.350  \,\, (\times 1) \,\,\, , \,\,\, 0  \,\, (\times 9) & ,
\end{array}
\eeq
which does not contain instabilities or flat directions. The masses of the vectors are
\beq
\label{vacuum8_vectors}
m^2 L^2 = 12 \,\, (\times 3) \,\,\, , \,\,\, \tfrac{2}{5}\, (21 \pm \sqrt{46})  \,\, (\times 3)   \,\,\, , \,\,\, 0 \,\, (\times 3) \ ,
\eeq
featuring the three massless vectors of the SO(3) residual symmetry. This solution is non-supersymmetric and perturbatively stable.


\section{Backgrounds with KK6$_{\parallel}$ (KKO6$_{\parallel}$) $\&$ D6$_{\parallel}$ (O6$_{\parallel}$)}
\label{app:both_KK6}

In this second appendix we present the detailed analysis of the nine M-theory backgrounds in section~\ref{sec:both_KK6} which include both KK6 (KKO6) $\rightarrow$ KK6$_{\parallel}$ (KKO6$_{\parallel}$) as well as KK6 (KKO6) $\rightarrow$ D6$_{\parallel}$ (O6$_{\parallel}$) sources. They all are this time compatible with the flux condition (\ref{fluxes_common_bothKK6}). As in the previous appendix, we present the associated background fluxes, M-theory superpotential, vacuum energy $V_{0}$, normalised mass spectra for scalars and vectors and preserved supersymmetry at each of the M-theory flux vacua.

\subsubsection*{Vacuum $9$}

The first of these solutions is induced by a flux background
\beq
\label{Fluxes_vacuum9}
a_{0} = -3 \, \lambda \,\,\,\, , \,\,\,\, a_{1} = b_{0} = c_{0} = 0 \,\,\,\, , \,\,\,\,  a_{2} = -c_{3}' = \frac{1}{2} \, \lambda  \,\,\,\, , \,\,\,\,  b_{1} = -d_{0} = \lambda  \ ,
\eeq
which implies a superpotential (\ref{Superpotential_Flux_MTheory_ISO}) of the form
\beq
\label{Superpotential_vacuum9}
W^{\textrm{(iso)}}_{\textrm{M-Theory}} = 3 \, \lambda \, (S \, T + S \,  U + \frac{1}{2} \, T^2 + \frac{1}{2} \, U^2 - 1)  \ .
\eeq
The value of the potential evaluated at this critical point is $V_{0}=-3 \lambda^2/8$. The set of scalar masses reads
\beq
\label{vacuum9_scalars}
m^2 L^2 = 18 \,\, (\times 1) \,\,\, , \,\,\, 10 \,\, (\times 6) \,\,\, , \,\,\, 4 \,\, (\times 6) \,\,\, , \,\,\, -2 \,\, (\times 18) \,\,\, , \,\,\, 0 \,\, (\times 7) \ ,
\eeq
not displaying instabilities but containing one flat direction. The set of vector masses is given by
\beq
\label{vacuum9_vectors}
m^2 L^2 = 12 \,\, (\times 3) \,\,\, , \,\,\, 6 \,\, (\times 3)  \,\,\, , \,\,\, 0 \,\, (\times 6) \ ,
\eeq
where the six massless vectors are associated to an enhancement of the SO(3) residual symmetry preserved by the STU models. Computing the gravitini mass matrix (\ref{gravitini_mass_matrix}) shows that this solution preserves $\,\mathcal{N}=3\,$ supersymmetry. Notice the similarities with the vacuum 1 discussed in the previous appendix.

\subsubsection*{Vacuum $10$}

This solution is generated from a flux background like (\ref{Fluxes_vacuum9}) after flipping the sign of the $a_{0}$ flux parameter. This is
\beq
\label{Fluxes_vacuum10}
a_{0} = 3 \, \lambda \,\,\,\, , \,\,\,\, a_{1} = b_{0} = c_{0} = 0 \,\,\,\, , \,\,\,\,  a_{2} = - c_{3}' =  \frac{1}{2} \, \lambda  \,\,\,\, , \,\,\,\,  b_{1} = - d_{0} =  \lambda \ ,
\eeq
and induces a superpotential
\beq
\label{Superpotential_vacuum10}
W^{\textrm{(iso)}}_{\textrm{M-Theory}} = 3 \, \lambda \, (S \, T + S \,  U + \frac{1}{2} \, T^2 + \frac{1}{2} \, U^2 + 1)  \ .
\eeq
At this critical point, the value of the vacuum energy is also $V_{0}=-3 \lambda^2/8$. Computing the spectrum of scalar masses we find
\beq
\label{vacuum10_scalars}
m^2 L^2 = 18 \,\, (\times 2) \,\,\, , \,\,\, 4 \,\, (\times 15) \,\,\, , \,\,\, -2 \,\, (\times 15)  \,\,\, , \,\,\, 0 \,\, (\times 6) \ ,
\eeq
not displaying instabilities and without flat directions. The mass spectrum for the vectors reads
\beq
\label{vacuum10_vectors}
m^2 L^2 = 12 \,\, (\times 3) \,\,\, , \,\,\, 6 \,\, (\times 3)  \,\,\, , \,\,\, 0 \,\, (\times 6) \ ,
\eeq
with six massless vectors being again associated to an enhancement of the SO(3) residual symmetry. This solution turns out to be non-supersymmetric and perturbatively stable.

\subsubsection*{Vacuum $11$}

This solution is obtained from the flux background
\beq
\label{Fluxes_vacuum11}
a_{0} = \frac{9}{5} \, \lambda \,\,\, , \,\,\, a_{1} = b_{0} = c_{0} = 0 \,\,\, , \,\,\,  a_{2} = -c_{3}'  = \frac{1}{2} \, \lambda  \,\,\,, \,\,\,  b_{1} =-  d_{0} =  \frac{1}{5} \, \lambda \ ,
\eeq
which induces the flux superpotential
\beq
\label{Superpotential_vacuum11}
W^{\textrm{(iso)}}_{\textrm{M-Theory}} = \frac{3}{5} \, \lambda \left( S T+S U+\frac{5}{2} T^2 +\frac{5}{2} U^2+ 3 \right)  \ .
\eeq
The value of the vacuum energy is $V_{0}=- 27 \lambda^2/200$. The scalar mass spectrum at this critical point consists of
\beq
\label{vacuum11_scalars}
\begin{array}{ccll}
m^2 L^2 &=& 10 \,\, (\times 1) \,\,\, , \,\,\, 18 \,\, (\times 1) \,\,\, , \,\,\, \frac{22}{9}  \,\, (\times 9) \,\,\, , \,\,\, \frac{70}{9}  \,\, (\times 9) \,\,\, , \,\,\, \frac{52}{9}  \,\, (\times 5)  & ,\\[2mm]
& & -\frac{20}{9} \,\, (\times 1) \,\,\, , \,\,\, \frac{10}{9}  \,\, (\times 5) \,\,\, , \,\,\, -\frac{8}{9}  \,\, (\times 1) \,\,\, , \,\,\, 0  \,\, (\times 6) & ,
\end{array}
\eeq
thus not containing instabilities nor flat directions. The set of vector masses is given by
\beq
\label{vacuum11_vectors}
m^2 L^2 = 12 \,\, (\times 3) \,\,\, , \,\,\, 6 \,\, (\times 3) \,\,\, , \,\,\,  0 \,\, (\times 6) \ ,
\eeq
showing six massless vectors associated to the residual symmetry. This solution preserves $\mathcal{N}=1$ supersymmetry.

\subsubsection*{Vacuum $12$}

This solution is related to that in (\ref{Fluxes_vacuum11}) again by a sign flip of the $a_{0}$ flux. The flux background reads
\beq
\label{Fluxes_vacuum12}
a_{0} = - \frac{9}{5} \, \lambda \,\,\, , \,\,\, a_{1} = b_{0} = c_{0} = 0 \,\,\, , \,\,\,  a_{2} = -c_{3}' = \frac{1}{2} \, \lambda  \,\,\,, \,\,\,  b_{1} =- d_{0} =  \frac{1}{5} \, \lambda \ ,
\eeq
and induces the moduli superpotential
\beq
\label{Superpotential_vacuum12}
W^{\textrm{(iso)}}_{\textrm{M-Theory}} = \frac{3}{5} \, \lambda \left( S T+S U+\frac{5}{2} T^2 +\frac{5}{2} U^2- 3 \right)  \ .
\eeq
As for the previous case, the vacuum energy is $V_{0}=- 27 \lambda^2/200$. The computation of the scalar masses gives
\beq
\label{vacuum12_scalars}
\begin{array}{ccll}
m^2 L^2 &=& \frac{190}{9}  \,\, (\times 5) \,\,\, , \,\,\,  18 \,\, (\times 1) \,\,\, , \,\,\, \frac{52}{9}  \,\, (\times 5) \,\,\, , \,\,\, \frac{22}{9}  \,\, (\times 9) \,\,\, , \,\,\, -\frac{20}{9}  \,\, (\times 10)  & ,\\[2mm]
& & -2 \,\, (\times 1) \,\,\, , \,\,\, \frac{10}{9}  \,\, (\times 1) \,\,\, , \,\,\, 0  \,\, (\times 6) & ,
\end{array}
\eeq
so it does not present instabilities or flat directions. The mass spectrum for the vectors consists of 
\beq
\label{vacuum12_vectors}
m^2 L^2 = 12 \,\, (\times 3) \,\,\, , \,\,\, 6 \,\, (\times 3) \,\,\, , \,\,\,  0 \,\, (\times 6) \ ,
\eeq
featuring the usual six massless vectors associated to the residual symmetry of these solutions. Supersymmetry is completely broken at this solution.

\subsubsection*{Vacuum $13$}

This solution corresponds to background fluxes of the form
\beq
\label{Fluxes_vacuum13}
a_{0} = -\frac{6}{5} \, \lambda \,\,\,\, , \,\,\,\, a_{1} = b_{0} = -c_{0} = \frac{3}{5} \, \lambda \,\,\,\, , \,\,\,\, a_{2} = -c_{3}' =\frac{1}{2} \, \lambda  \,\,\,\, , \,\,\,\,  b_{1} = - d_{0} =  \frac{1}{5} \, \lambda \ ,
\eeq
and induces the flux superpotential
\beq
\label{Superpotential_vacuum13}
W^{\textrm{(iso)}}_{\textrm{M-Theory}} = \frac{3}{5} \, \lambda \left( -S -3 T -3\, U + S T+S U+ \frac{5}{2}\, T^2  + \frac{5}{2}\, U^2 - 2 \right)  \ .
\eeq
It has a vacuum energy $V_{0}=- 9 \lambda^2/80$. The mass spectrum for the scalars is given by
\beq
\label{vacuum13_scalars}
\begin{array}{ccll}
m^2 L^2 &=& \frac{76}{3} \,\, (\times 5) \,\,\, , \,\,\, 18 \,\, (\times 1) \,\,\, , \,\,\,  \frac{22}{3}  \,\, (\times 9) \,\,\, , \,\,\, \frac{16}{3}  \,\, (\times 5) \,\,\, , \,\,\, -\frac{8}{3}  \,\, (\times 9) \,\,\, , \,\,\, 6  \,\, (\times 1) & ,\\[2mm]
& & \frac{4}{3}  \,\, (\times 1) \,\,\, , \,\,\, -\frac{2}{3}  \,\, (\times 1) \,\,\, , \,\,\, 0  \,\, (\times 6) & ,
\end{array}
\eeq
which contains instabilities (modes with $m^2 L^2=-\frac{8}{3}$) and has no flat directions. The vector masses are
\beq
\label{vacuum13_vectors}
m^2 L^2 = 12 \,\, (\times 6)  \,\,\, , \,\,\, 0 \,\, (\times 6) \ ,
\eeq
containing six massless vectors associated to the residual symmetry. The gravitini mass matrix shows that this solution is non-supersymmetric.

\subsubsection*{Vacuum $14$}

This solution is generated by a flux background given by
\beq
\label{Fluxes_vacuum14}
\begin{array}{lc}
a_{0} = -\lambda  \,\,\,,\,\,\,  a_{1} =c_{0}=  -\tfrac{1}{2} \, \sqrt{ \tfrac{1}{6} \,(31+3\,\sqrt{57}) }\, \lambda   \,\,\,,\,\,\,    b_{0} = 0  \,\,\,\, , \,\,\,\, a_{2} = -c_{3}' =\frac{1}{2} \, \lambda   & , \\[2mm]
b_{1} = -d_{0} = \tfrac{1}{12} \, (9+\sqrt{57}) \, \lambda & ,
\end{array}
\eeq
which is similar to (\ref{Fluxes_vacuum6}). The associated superpotential also lacks the linear term on the dilaton modulus $S$. The vacuum energy is $V_{0}=- \tfrac{1}{64} \, (13 + \sqrt{57})\, \lambda^2$. We find the scalar mass spectrum
\beq
\label{vacuum14_scalars}
\begin{array}{ccll}
m^2 L^2 &=& 21.308 \,\, (\times 1) \,\,\, , \,\,\, 18 \,\, (\times 1) \,\,\, , \,\,\,  13.328  \,\, (\times 1) \,\,\, , \,\,\, 13.771  \,\, (\times 1)  & ,\\[2mm]
& & 12.942 \,\, (\times 5) \,\,\, , \,\,\, 12.036 \,\, (\times 5)  \,\,\, , \,\,\,  9.221 \,\, (\times 1) \,\,\, , \,\,\, -5.643  \,\, (\times 1)  & , \\[2mm]
& & -2.693  \,\, (\times 5) \,\,\, , \,\,\, 2.336  \,\, (\times 3) \,\,\, , \,\,\, -0.942  \,\, (\times 5) \,\,\, , \,\,\, 0  \,\, (\times 9) & ,
\end{array}
\eeq
which displays tachyons violating the BF bound and has no flat directions. The mass spectrum for the vectors reads
\beq
\label{vacuum14_vectors}
m^2 L^2 = \frac{1}{7} \left(71+\sqrt{57} \pm \sqrt{1186-74 \sqrt{57}}\right)  \,\, (\times 3)   \,\,\, , \,\,\, \tfrac{1}{7}\, (71 + \sqrt{57})  \,\, (\times 3)   \,\,\, , \,\,\, 0 \,\, (\times 3) \ ,
\eeq
containing the three massless vectors of the SO(3) residual symmetry. This is a non-supersymmetric and unstable solution.

\subsubsection*{Vacuum $15$}

This solution is induced by a family of fluxes of the form
\beq
\label{Fluxes_vacuum15}
\begin{array}{lc}
a_{0} = -\lambda  \,\,\,,\,\,\,  a_{1} =c_{0}=  -\tfrac{1}{2} \, \sqrt{ \tfrac{1}{6} \,(31-3\,\sqrt{57}) }\, \lambda   \,\,\,,\,\,\,    b_{0} = 0  \,\,\,\, , \,\,\,\, a_{2} = -c_{3}' =\frac{1}{2} \, \lambda   & , \\[2mm]
b_{1} = -d_{0} = \tfrac{1}{12} \, (9-\sqrt{57}) \, \lambda & ,
\end{array}
\eeq
which is similar to (\ref{Fluxes_vacuum14}) and produces a vacuum energy $V_{0}=- \tfrac{1}{64} \, (13 - \sqrt{57})\, \lambda^2$. The spectrum of scalar masses is given by
\beq
\label{vacuum15_scalars}
\begin{array}{ccll}
m^2 L^2 &=& 31.235 \,\, (\times 5) \,\,\, , \,\,\, 18 \,\, (\times 1) \,\,\, , \,\,\,  13.936  \,\, (\times 5) \,\,\, , \,\,\, 11.458  \,\, (\times 1)  & ,\\[2mm]
& & 8.807 \,\, (\times 3) \,\,\, , \,\,\,  7.064 \,\, (\times 1)  \,\,\, , \,\,\,  3.994 \,\, (\times 5) \,\,\, , \,\,\, -2.265  \,\, (\times 1)  & , \\[2mm]
& & -3.486  \,\, (\times 1) \,\,\, , \,\,\, -1.936  \,\, (\times 5) \,\,\, , \,\,\, 0.386  \,\, (\times 1) \,\,\, , \,\,\, 0  \,\, (\times 9) & ,
\end{array}
\eeq
showing two tachyon masses that violate the BF bound. The solution is unstable and does not contain flat directions. Computing the vector masses, they are given by
\beq
\label{vacuum15_vectors}
m^2 L^2 = \frac{1}{7} \left(71-\sqrt{57}\pm\sqrt{1186+74 \sqrt{57}}\right)  \,\, (\times 3)   \,\,\, , \,\,\, \tfrac{1}{7}\, (71 - \sqrt{57})  \,\, (\times 3)   \,\,\, , \,\,\, 0 \,\, (\times 3) \ ,
\eeq
then resembling those in (\ref{vacuum14_vectors}) and also including the three SO(3) massless vectors. Looking at the gravitini mass matrix, one finds that this solution is non-supersymmetric.

\subsubsection*{Vacuum $16$}

The flux background associated to this solution turns out to be quite involved
\beq
\label{Fluxes_vacuum16}
\begin{array}{llllllll}
a_{0} = 1.546 \, \lambda  & \,\,,\,\, &  a_{1} = 0.712 \, \lambda  & \,\,,\,\, &   c_{0}=  - 0.542 \, \lambda   & \,\,,\,\, &    b_{0} = -2.632 \, \lambda   & , \\[2mm]
a_{2} = 2.231 \, \lambda    & \,\,,\,\, &   c_{3}' =-0.112 \, \lambda   & \,\,,\,\, &  b_{1} =  1.626 \, \lambda  & \,\,,\,\, &   d_{0} = -0.364 \, \lambda & .
\end{array}
\eeq
Evaluated at this solution, the vacuum energy is $V_{0}=-0.294\, \lambda^2$. The spectrum of scalar masses reads
\beq
\label{vacuum16_scalars}
\begin{array}{ccll}
m^2 L^2 &=& 66.413 \,\, (\times 1) \,\,\, , \,\,\, 18 \,\, (\times 1) \,\,\, , \,\,\,  24.061  \,\, (\times 1) \,\,\, , \,\,\, 21.246  \,\, (\times 5)  & ,\\[2mm]
& & 12.341 \,\, (\times 3) \,\,\, , \,\,\, 7.887 \,\, (\times 1)  \,\,\, , \,\,\,  4.122 \,\, (\times 5) \,\,\, , \,\,\, -2.908  \,\, (\times 1)  & , \\[2mm]
& & 2.040  \,\, (\times 5) \,\,\, , \,\,\, -1.857  \,\, (\times 5) \,\,\, , \,\,\, 1.089  \,\, (\times 1) \,\,\, , \,\,\, 0  \,\, (\times 9) & ,
\end{array}
\eeq
and contains one tachyon violating the BF bound and no flat directions. The set of vectors masses is given by
\beq
\label{vacuum16_vectors}
m^2 L^2 = 23.764 \,\, (\times 3)   \,\,\, , \,\,\, 18.900 \,\, (\times 3)   \,\,\, , \,\,\, 3.920 \,\, (\times 3)   \,\,\, , \,\,\, 0 \,\, (\times 3) \ ,
\eeq
showing the three massless vectors of the SO(3) symmetry. This solution is a non-supersymmetric critical point. We want to mention that there is a companion flux background obtained by exchanging $a_{1} \leftrightarrow -c_{0}$, $a_{2} \leftrightarrow -c_{3}'$ and $b_{1} \leftrightarrow -d_{0}$ which produces the same mass spectra (\ref{vacuum16_scalars}) and  (\ref{vacuum16_vectors}) and is also non-supersymmetric.

\subsubsection*{Vacuum $17$}

This is the last solution of the STU model we explore in this paper. The associated flux background reads
\beq
\label{Fluxes_vacuum17}
\begin{array}{llllllll}
a_{0} = -0.683 \, \lambda  & \,\,,\,\, &  a_{1} =-0.837 \, \lambda  & \,\,,\,\, &   c_{0}=  -0.034\, \lambda   & \,\,,\,\, &    b_{0} = -0.252 \, \lambda   & , \\[2mm]
a_{2} = 0.330 \, \lambda    & \,\,,\,\, &   c_{3}' = -0.757 \, \lambda   & \,\,,\,\, &  b_{1} =  0.073 \, \lambda  & \,\,,\,\, &   d_{0} = -0.111 \, \lambda & ,
\end{array}
\eeq
and produces a vacuum energy $V_{0}=-0.065\, \lambda^2$. The scalar masses are given by
\beq
\label{vacuum17_scalars}
\begin{array}{ccll}
m^2 L^2 &=& 40.404 \,\, (\times 5) \,\,\, , \,\,\, 18 \,\, (\times 1) \,\,\, , \,\,\,  19.899  \,\, (\times 1) \,\,\, , \,\,\, 16.974 \,\, (\times 3)  & ,\\[2mm]
& & 15.001 \,\, (\times 5) \,\,\, , \,\,\, 11.576 \,\, (\times 5)  \,\,\, , \,\,\,  7.810 \,\, (\times 1) \,\,\, , \,\,\, 6.556 \,\, (\times 1)  & , \\[2mm]
& & -3.815 \,\, (\times 1) \,\,\, , \,\,\, -2.102  \,\, (\times 5) \,\,\, , \,\,\,  0.201  \,\, (\times 1) \,\,\, , \,\,\, 0  \,\, (\times 9) & .
\end{array}
\eeq
There is one tachyon violating the BF bound and no flat directions. The spectrum of vector masses is
\beq
\label{vacuum17_vectors}
m^2 L^2 = 17.563 \,\, (\times 3)   \,\,\, , \,\,\, 15.305 \,\, (\times 3)   \,\,\, , \,\,\, 3.318 \,\, (\times 3)   \,\,\, , \,\,\, 0 \,\, (\times 3) \ ,
\eeq
containing the three SO(3) massless vectors. Computing the gravitino mass matrix shows that this solution is non-supersymmetric. Finally, there is also a companion background if exchanging $a_{1} \leftrightarrow -c_{0}$, $a_{2} \leftrightarrow -c_{3}'$ and $b_{1} \leftrightarrow -d_{0}$ which produces the same spectra (\ref{vacuum17_scalars}) and  (\ref{vacuum17_vectors}) and is also non-supersymmetric.

%
%

\small
\bibliography{references}

\providecommand{\href}[2]{#2}\begingroup\raggedright\begin{thebibliography}{10}

\bibitem{Scherk:1979zr}
J.~Scherk and J.~H. Schwarz, ``{How to Get Masses from Extra Dimensions},''
\href{http://dx.doi.org/10.1016/0550-3213(79)90592-3}{{\em Nucl.Phys.} {\bf
  B153} (1979)  61--88}.

\bibitem{deWit:2007mt}
B.~de~Wit, H.~Samtleben, and M.~Trigiante, ``{The maximal D=4
  supergravities},''
  \href{http://dx.doi.org/10.1088/1126-6708/2007/06/049}{{\em JHEP} {\bf 06}
  (2007)  049}, \href{http://arxiv.org/abs/0705.2101}{{\tt arXiv:0705.2101
  [hep-th]}}.

\bibitem{Schon:2006kz}
J.~Schon and M.~Weidner, ``{Gauged N=4 supergravities},''
  \href{http://dx.doi.org/10.1088/1126-6708/2006/05/034}{{\em JHEP} {\bf 0605}
  (2006)  034},
\href{http://arxiv.org/abs/hep-th/0602024}{{\tt arXiv:hep-th/0602024
  [hep-th]}}.

\bibitem{Derendinger:2004jn}
J.-P. Derendinger, C.~Kounnas, P.~M. Petropoulos, and F.~Zwirner,
  ``{Superpotentials in IIA compactifications with general fluxes},''
  \href{http://dx.doi.org/10.1016/j.nuclphysb.2005.02.038}{{\em Nucl.Phys.}
  {\bf B715} (2005)  211--233},
\href{http://arxiv.org/abs/hep-th/0411276}{{\tt arXiv:hep-th/0411276
  [hep-th]}}.

\bibitem{Kachru:2003aw}
S.~Kachru, R.~Kallosh, A.~D. Linde, and S.~P. Trivedi, ``{De Sitter vacua in
  string theory},'' \href{http://dx.doi.org/10.1103/PhysRevD.68.046005}{{\em
  Phys.Rev.} {\bf D68} (2003)  046005},
\href{http://arxiv.org/abs/hep-th/0301240}{{\tt arXiv:hep-th/0301240
  [hep-th]}}.

\bibitem{Villadoro:2007yq}
G.~Villadoro and F.~Zwirner, ``{Beyond Twisted Tori},''
  \href{http://dx.doi.org/10.1016/j.physletb.2007.07.002}{{\em Phys.Lett.} {\bf
  B652} (2007)  118--123},
\href{http://arxiv.org/abs/0706.3049}{{\tt arXiv:0706.3049 [hep-th]}}.

\bibitem{Andriot:2014uda}
D.~Andriot and A.~Betz, ``{NS-branes, source corrected Bianchi identities, and
  more on backgrounds with non-geometric fluxes},''
  \href{http://dx.doi.org/10.1007/JHEP07(2014)059}{{\em JHEP} {\bf 1407} (2014)
   059},
\href{http://arxiv.org/abs/1402.5972}{{\tt arXiv:1402.5972 [hep-th]}}.

\bibitem{Dall'Agata:2009gv}
G.~Dall'Agata, G.~Villadoro, and F.~Zwirner, ``{Type-IIA flux compactifications
  and N=4 gauged supergravities},''
  \href{http://dx.doi.org/10.1088/1126-6708/2009/08/018}{{\em JHEP} {\bf 0908}
  (2009)  018},
\href{http://arxiv.org/abs/0906.0370}{{\tt arXiv:0906.0370 [hep-th]}}.

\bibitem{Dibitetto:2011gm}
G.~Dibitetto, A.~Guarino, and D.~Roest, ``{Charting the landscape of N=4 flux
  compactifications},'' \href{http://dx.doi.org/10.1007/JHEP03(2011)137}{{\em
  JHEP} {\bf 1103} (2011)  137}, \href{http://arxiv.org/abs/1102.0239}{{\tt
  arXiv:1102.0239 [hep-th]}}.

\bibitem{Romans:1985tz}
L.~Romans, ``{Massive N=2a Supergravity in Ten-Dimensions},''
\href{http://dx.doi.org/10.1016/0370-2693(86)90375-8}{{\em Phys.Lett.} {\bf
  B169} (1986)  374}.

\bibitem{House:2004pm}
T.~House and A.~Micu, ``{M-Theory compactifications on manifolds with G(2)
  structure},'' \href{http://dx.doi.org/10.1088/0264-9381/22/9/016}{{\em
  Class.Quant.Grav.} {\bf 22} (2005)  1709--1738},
\href{http://arxiv.org/abs/hep-th/0412006}{{\tt arXiv:hep-th/0412006
  [hep-th]}}.

\bibitem{DallAgata:2005fm}
G.~Dall'Agata and N.~Prezas, ``{Scherk-Schwarz reduction of M-theory on
  G2-manifolds with fluxes},''
  \href{http://dx.doi.org/10.1088/1126-6708/2005/10/103}{{\em JHEP} {\bf 0510}
  (2005)  103},
\href{http://arxiv.org/abs/hep-th/0509052}{{\tt arXiv:hep-th/0509052
  [hep-th]}}.

\bibitem{Hull:2006tp}
C.~Hull and R.~Reid-Edwards, ``{Flux compactifications of M-theory on twisted
  Tori},'' \href{http://dx.doi.org/10.1088/1126-6708/2006/10/086}{{\em JHEP}
  {\bf 0610} (2006)  086},
\href{http://arxiv.org/abs/hep-th/0603094}{{\tt arXiv:hep-th/0603094
  [hep-th]}}.

\bibitem{Looyestijn:2010pb}
H.~Looyestijn, E.~Plauschinn, and S.~Vandoren, ``{New potentials from
  Scherk-Schwarz reductions},''
  \href{http://dx.doi.org/10.1007/JHEP12(2010)016}{{\em JHEP} {\bf 1012} (2010)
   016},
\href{http://arxiv.org/abs/1008.4286}{{\tt arXiv:1008.4286 [hep-th]}}.

\bibitem{Castellani:1983tc}
L.~Castellani and L.~Romans, ``{$N=3$ and $N=1$ Supersymmetry in a New Class of
  Solutions for $d=11$ Supergravity},''
\href{http://dx.doi.org/10.1016/0550-3213(84)90343-2}{{\em Nucl.Phys.} {\bf
  B238} (1984)  683--701}.

\bibitem{Castellani:1983yg}
L.~Castellani, L.~Romans, and N.~Warner, ``{A Classification of Compactifying
  Solutions for $d=11$ Supergravity},''
\href{http://dx.doi.org/10.1016/0550-3213(84)90055-5}{{\em Nucl.Phys.} {\bf
  B241} (1984)  429}.

\bibitem{Cassani:2011fu}
D.~Cassani and P.~Koerber, ``{Tri-Sasakian consistent reduction},''
  \href{http://dx.doi.org/10.1007/JHEP01(2012)086}{{\em JHEP} {\bf 1201} (2012)
   086},
\href{http://arxiv.org/abs/1110.5327}{{\tt arXiv:1110.5327 [hep-th]}}.

\bibitem{Cassani:2012pj}
D.~Cassani, P.~Koerber, and O.~Varela, ``{All homogeneous N=2 M-theory
  truncations with supersymmetric AdS4 vacua},''
  \href{http://dx.doi.org/10.1007/JHEP11(2012)173}{{\em JHEP} {\bf 1211} (2012)
   173},
\href{http://arxiv.org/abs/1208.1262}{{\tt arXiv:1208.1262 [hep-th]}}.

\bibitem{Aldazabal:2006up}
G.~Aldazabal, P.~G. Camara, A.~Font, and L.~Ibanez, ``{More dual fluxes and
  moduli fixing},'' \href{http://dx.doi.org/10.1088/1126-6708/2006/05/070}{{\em
  JHEP} {\bf 0605} (2006)  070},
\href{http://arxiv.org/abs/hep-th/0602089}{{\tt arXiv:hep-th/0602089
  [hep-th]}}.

\bibitem{Hull:1998vy}
C.~Hull, ``{Massive string theories from M theory and F theory},''
  \href{http://dx.doi.org/10.1088/1126-6708/1998/11/027}{{\em JHEP} {\bf 9811}
  (1998)  027},
\href{http://arxiv.org/abs/hep-th/9811021}{{\tt arXiv:hep-th/9811021
  [hep-th]}}.

\bibitem{Shelton:2005cf}
J.~Shelton, W.~Taylor, and B.~Wecht, ``{Nongeometric flux compactifications},''
  \href{http://dx.doi.org/10.1088/1126-6708/2005/10/085}{{\em JHEP} {\bf 0510}
  (2005)  085},
\href{http://arxiv.org/abs/hep-th/0508133}{{\tt arXiv:hep-th/0508133
  [hep-th]}}.

\bibitem{Grana:2013ila}
M.~Gra\~na, R.~Minasian, H.~Triendl, and T.~Van~Riet, ``{Quantization problem
  in Scherk-Schwarz compactifications},''
  \href{http://dx.doi.org/10.1103/PhysRevD.88.085018}{{\em Phys.Rev.} {\bf D88}
  (2013) no.~8, 085018},
\href{http://arxiv.org/abs/1305.0785v3}{{\tt arXiv:1305.0785v3 [hep-th]}}.

\bibitem{Beasley:2002db}
C.~Beasley and E.~Witten, ``{A Note on fluxes and superpotentials in M theory
  compactifications on manifolds of G(2) holonomy},''
  \href{http://dx.doi.org/10.1088/1126-6708/2002/07/046}{{\em JHEP} {\bf 0207}
  (2002)  046},
\href{http://arxiv.org/abs/hep-th/0203061}{{\tt arXiv:hep-th/0203061
  [hep-th]}}.

\bibitem{Gukov:1999gr}
S.~Gukov, ``{Solitons, superpotentials and calibrations},''
  \href{http://dx.doi.org/10.1016/S0550-3213(00)00053-5}{{\em Nucl.Phys.} {\bf
  B574} (2000)  169--188},
\href{http://arxiv.org/abs/hep-th/9911011}{{\tt arXiv:hep-th/9911011
  [hep-th]}}.

\bibitem{Acharya:2000ps}
B.~S. Acharya and B.~J. Spence, ``{Flux, supersymmetry and M theory on seven
  manifolds},''
\href{http://arxiv.org/abs/hep-th/0007213}{{\tt arXiv:hep-th/0007213
  [hep-th]}}.

\bibitem{Font:2008vd}
A.~Font, A.~Guarino, and J.~M. Moreno, ``{Algebras and non-geometric flux
  vacua},'' \href{http://dx.doi.org/10.1088/1126-6708/2008/12/050}{{\em JHEP}
  {\bf 0812} (2008)  050},
\href{http://arxiv.org/abs/0809.3748}{{\tt arXiv:0809.3748 [hep-th]}}.

\bibitem{Friedrich:2002}
T.~Friedrich, ``{On types of non-integrable geometries},''
  \href{http://arxiv.org/abs/eprint arXiv:math/0205149}{{\tt eprint
  arXiv:math/0205149}}.

\bibitem{Bryant:2003}
R.~L. Bryant, ``{Some remarks on G2-structures},''
  \href{http://dx.doi.org/10.1088/1126-6708/2006/05/034}{{\em Proceeding of
  Gokova Geometry-Topology Conference} (2005)  },
  \href{http://arxiv.org/abs/arXiv:math/0305124}{{\tt arXiv:math/0305124}}.

\bibitem{Held:2011uz}
J.~Held, ``{BPS-like potential for compactifications of heterotic M-theory?},''
  \href{http://dx.doi.org/10.1007/JHEP10(2011)136}{{\em JHEP} {\bf 1110} (2011)
   136},
\href{http://arxiv.org/abs/1109.1974}{{\tt arXiv:1109.1974 [hep-th]}}.

\bibitem{Villadoro:2005cu}
G.~Villadoro and F.~Zwirner, ``{N=1 effective potential from dual type-IIA
  D6/O6 orientifolds with general fluxes},''
  \href{http://dx.doi.org/10.1088/1126-6708/2005/06/047}{{\em JHEP} {\bf 0506}
  (2005)  047},
\href{http://arxiv.org/abs/hep-th/0503169}{{\tt arXiv:hep-th/0503169
  [hep-th]}}.

\bibitem{Dibitetto:2012ia}
G.~Dibitetto, A.~Guarino, and D.~Roest, ``{Exceptional Flux
  Compactifications},'' \href{http://dx.doi.org/10.1007/JHEP05(2012)056}{{\em
  JHEP} {\bf 1205} (2012)  056},
\href{http://arxiv.org/abs/1202.0770}{{\tt arXiv:1202.0770 [hep-th]}}.

\bibitem{Dall'Agata:2005ff}
G.~Dall'Agata and S.~Ferrara, ``{Gauged supergravity algebras from twisted tori
  compactifications with fluxes},''
  \href{http://dx.doi.org/10.1016/j.nuclphysb.2005.03.039}{{\em Nucl.Phys.}
  {\bf B717} (2005)  223--245},
\href{http://arxiv.org/abs/hep-th/0502066}{{\tt arXiv:hep-th/0502066
  [hep-th]}}.

\bibitem{Dall'Agata:2005mj}
G.~Dall'Agata, R.~D'Auria, and S.~Ferrara, ``{Compactifications on twisted tori
  with fluxes and free differential algebras},''
  \href{http://dx.doi.org/10.1016/j.physletb.2005.04.005}{{\em Phys.Lett.} {\bf
  B619} (2005)  149--154},
\href{http://arxiv.org/abs/hep-th/0503122}{{\tt arXiv:hep-th/0503122
  [hep-th]}}.

\bibitem{deRoo:1985jh}
M.~de~Roo and P.~Wagemans, ``{Gauge Matter Coupling in $N=4$ Supergravity},''
\href{http://dx.doi.org/10.1016/0550-3213(85)90509-7}{{\em Nucl.Phys.} {\bf
  B262} (1985)  644}.

\bibitem{Dibitetto:2011eu}
G.~Dibitetto, A.~Guarino, and D.~Roest, ``{How to halve maximal
  supergravity},'' \href{http://dx.doi.org/10.1007/JHEP06(2011)030}{{\em JHEP}
  {\bf 1106} (2011)  030},
\href{http://arxiv.org/abs/1104.3587}{{\tt arXiv:1104.3587 [hep-th]}}.

\bibitem{Hertzberg:2007wc}
M.~P. Hertzberg, S.~Kachru, W.~Taylor, and M.~Tegmark, ``{Inflationary
  Constraints on Type IIA String Theory},''
  \href{http://dx.doi.org/10.1088/1126-6708/2007/12/095}{{\em JHEP} {\bf 0712}
  (2007)  095},
\href{http://arxiv.org/abs/0711.2512}{{\tt arXiv:0711.2512 [hep-th]}}.

\bibitem{Silverstein:2007ac}
E.~Silverstein, ``{Simple de Sitter Solutions},''
  \href{http://dx.doi.org/10.1103/PhysRevD.77.106006}{{\em Phys.Rev.} {\bf D77}
  (2008)  106006},
\href{http://arxiv.org/abs/0712.1196}{{\tt arXiv:0712.1196 [hep-th]}}.

\bibitem{Haque:2008jz}
S.~S. Haque, G.~Shiu, B.~Underwood, and T.~Van~Riet, ``{Minimal simple de
  Sitter solutions},'' \href{http://dx.doi.org/10.1103/PhysRevD.79.086005}{{\em
  Phys.Rev.} {\bf D79} (2009)  086005},
\href{http://arxiv.org/abs/0810.5328}{{\tt arXiv:0810.5328 [hep-th]}}.

\bibitem{Caviezel:2008tf}
C.~Caviezel, P.~Koerber, S.~Kors, D.~Lust, T.~Wrase, {\em et al.}, ``{On the
  Cosmology of Type IIA Compactifications on SU(3)-structure Manifolds},''
  \href{http://dx.doi.org/10.1088/1126-6708/2009/04/010}{{\em JHEP} {\bf 0904}
  (2009)  010},
\href{http://arxiv.org/abs/0812.3551}{{\tt arXiv:0812.3551 [hep-th]}}.

\bibitem{Danielsson:2009ff}
U.~H. Danielsson, S.~S. Haque, G.~Shiu, and T.~Van~Riet, ``{Towards Classical
  de Sitter Solutions in String Theory},''
  \href{http://dx.doi.org/10.1088/1126-6708/2009/09/114}{{\em JHEP} {\bf 0909}
  (2009)  114},
\href{http://arxiv.org/abs/0907.2041}{{\tt arXiv:0907.2041 [hep-th]}}.

\bibitem{Dibitetto:2014sfa}
G.~Dibitetto, A.~Guarino, and D.~Roest, ``{Lobotomy of Flux
  Compactifications},'' \href{http://dx.doi.org/10.1007/JHEP05(2014)067}{{\em
  JHEP} {\bf 1405} (2014)  067},
\href{http://arxiv.org/abs/1402.4478}{{\tt arXiv:1402.4478 [hep-th]}}.

\bibitem{DGPS}
W.~Decker, G.-M. Greuel, G.~Pfister, and H.~Sch\"onemann, ``{\sc Singular}
  {3-1-5} --- {A} computer algebra system for polynomial computations.''
  Http://www.singular.uni-kl.de, 2012.

\bibitem{Borghese:2010ei}
A.~Borghese and D.~Roest, ``{Metastable supersymmetry breaking in extended
  supergravity},'' \href{http://dx.doi.org/10.1007/JHEP05(2011)102}{{\em JHEP}
  {\bf 1105} (2011)  102},
\href{http://arxiv.org/abs/1012.3736}{{\tt arXiv:1012.3736 [hep-th]}}.

\bibitem{Camara:2005dc}
P.~G. Camara, A.~Font, and L.~Ibanez, ``{Fluxes, moduli fixing and MSSM-like
  vacua in a simple IIA orientifold},''
  \href{http://dx.doi.org/10.1088/1126-6708/2005/09/013}{{\em JHEP} {\bf 0509}
  (2005)  013},
\href{http://arxiv.org/abs/hep-th/0506066}{{\tt arXiv:hep-th/0506066
  [hep-th]}}.

\bibitem{Dibitetto&Danielsson}
G.~Dibitetto. {\em Private communication}  .

\bibitem{Freund:1980xh}
P.~G. Freund and M.~A. Rubin, ``{Dynamics of Dimensional Reduction},''
\href{http://dx.doi.org/10.1016/0370-2693(80)90590-0}{{\em Phys.Lett.} {\bf
  B97} (1980)  233--235}.

\bibitem{Dabholkar:2002sy}
A.~Dabholkar and C.~Hull, ``{Duality twists, orbifolds, and fluxes},''
  \href{http://dx.doi.org/10.1088/1126-6708/2003/09/054}{{\em JHEP} {\bf 0309}
  (2003)  054},
\href{http://arxiv.org/abs/hep-th/0210209}{{\tt arXiv:hep-th/0210209
  [hep-th]}}.

\bibitem{Dabholkar:2005ve}
A.~Dabholkar and C.~Hull, ``{Generalised T-duality and non-geometric
  backgrounds},'' \href{http://dx.doi.org/10.1088/1126-6708/2006/05/009}{{\em
  JHEP} {\bf 0605} (2006)  009},
\href{http://arxiv.org/abs/hep-th/0512005}{{\tt arXiv:hep-th/0512005
  [hep-th]}}.

\bibitem{Condeescu:2013yma}
C.~Condeescu, I.~Florakis, C.~Kounnas, and D.~Lüst, ``{Gauged supergravities
  and non-geometric Q/R-fluxes from asymmetric orbifold CFT`s},''
  \href{http://dx.doi.org/10.1007/JHEP10(2013)057}{{\em JHEP} {\bf 1310} (2013)
   057},
\href{http://arxiv.org/abs/1307.0999}{{\tt arXiv:1307.0999 [hep-th]}}.

\bibitem{Breitenlohner:1982bm}
P.~Breitenlohner and D.~Z. Freedman, ``{Positive Energy in anti-De Sitter
  Backgrounds and Gauged Extended Supergravity},''
\href{http://dx.doi.org/10.1016/0370-2693(82)90643-8}{{\em Phys.Lett.} {\bf
  B115} (1982)  197}.

\end{thebibliography}\endgroup
\bibliographystyle{utphys}
\end{document}